\documentstyle[aps,12pt]{revtex}

\widetext
\draft
\tighten
\begin{document}

\preprint{EFUAZ FT-96-32}

\title{Essay on the Non-Maxwellian Theories of
Electromagnetism\thanks{Submitted to ``Physics Essays"}}

\author{{\bf Valeri V. Dvoeglazov}}

\address {Escuela de F\'{\i}sica, Universidad Aut\'onoma de Zacatecas \\
Antonio Doval\'{\i} Jaime\, s/n, Zacatecas 98068, ZAC., M\'exico\\
Internet address:  VALERI@CANTERA.REDUAZ.MX
}

\date{Received \qquad\qquad, 1996}

\maketitle

\baselineskip13pt

\medskip

\begin{abstract}
In the first part of this  paper we review several formalisms which
give alternative ways for describing the light. They are:
the formalism `baroque' and the Majorana-Oppenheimer form of
electrodynamics, the Sachs' theory of Elementary Matter, the
Dirac-Fock-Podol'sky model, its development by Staruszkiewicz,
the Evans-Vigier ${\bf B}^{(3)}$ field, the theory with an invariant
evolution parameter of Horwitz, the analysis of the
action-at-a-distance concept, presented recently by Chubykalo and
Smirnov-Rueda, and the analysis of the claimed `longitudity' of the
antisymmetric tensor field after quantization.  The second part is devoted
to the discussion of the Weinberg formalism and its recent development by
Ahluwalia and myself.
\end{abstract}

\medskip

\section{Historical Notes}

The Maxwell's electromagnetic theory perfectly describes many observed
phenomena. The accuracy in predictions of the quantum electrodynamics
is without precedents~\cite{DVOQED}. They are widely accepted as the only
tools to deal with electromagnetic phenomena.  Other modern field theories
have been built on the basis of the similar principles to deal with weak,
strong and gravitational interactions.  Nevertheless,  many scientists
felt some unsatisfactions with both these theories  since
almost their appearance, see, {\it e.g.}, refs.~\cite{CRITICS} and
refs.~\cite{DIRAC,LANDAU,PAULI}.  In the preface to the Dover edition
of his book~\cite{BARUT2} A. Barut writes (1979): ``Electrodynamics and
the classical theory of fields remain very much alive and continue to be
the source of inspiration for much of the modern research work in new
physical theories"  and in the preface to the first edition he said
about shortcomings in the modern quantum field theory. They are well known.
Furthermore, in spite of much expectation in the sixties and the seventies
after the proposal of the Glashow-Salam-Weinberg model and the quantum
chromodynamics, elaboration of the unified field theory, based on the
gauge principle, has come across with serious difficulties.  In the end of
the nineties there are a lot of experiments in our disposition, which do
not find satisfactory explanations on the basis of the standard model.
First of all, one can single out the following ones:  the LANL neutrino
oscillation experiment; the atmospheric neutrino anomaly, the solar
neutrino puzzle (all of the above-mentioned imply existence of the
neutrino mass); the tensor coupling in decays of $\pi^-$ and $K^+$ mesons;
the dark matter problem; the observed periodicity of the number
distribution of galaxies, and the `spin crisis' in QCD.  Furthemore,
experiments and observations concerning with superluminal phenomena:
negative mass-square neutrinos, tunelling photons, ``X-shaped waves" and
superluminal expansions in quasars and in {\it galactic} objects.

In the meantime almost since the proposal of the
Lorentz-Poincar\`e-Einstein  theory of relativity~\cite{EINST} and
the development of the mathematical formalism of the Poincar\`e
group~\cite{WIGNER} several physicists (including A.  Einstein, W. Pauli
and M. Sachs) thought that in order to build a reliable theory, based on
relativistic ideas, one must utilize the irreducible representations
of the underlying symmetry group --- the Poincar\`e group of special
relativity, {\it i.e.} to build it on the first principles.  Considerable
efforts in this direction have been recently undertaken  by M.  Evans.
Since the prediction and the discovery of an additional phase-free
variable, the spin, which all observed fundamental particles have, to find
its classical analogue and to relate it with known fields and/or
space-time structures (perhaps, in higher dimensions) was one of the
important tasks of physicists. We can say now that this has been done (see
papers and books of M. Evans and what is below). In the end of this
introductory part we note that while the `Ultimate' Theory has not yet
been proposed recent papers of D.
V. Ahluwalia, M. W. Evans, E.  Recami and several other works provide a
sufficiently clear way to this goal.

\smallskip

We deal below with the historical development, with the ideas which could
be useful to proceed further.

\smallskip

{\it $E=0$ solutions.} First of all, I would like to mention the problem
with existence of `acausal' solutions of relativistic wave equations of the
first order.  In ref.~\cite{DVA01} and then in~\cite{DVA02} the author
found that massless equations of the form\footnote{Here and  below in this
historical section we try to keep the notation and the metric of
original papers.}
\begin{mathletters}
\begin{eqnarray}
\left ({\bf J}\cdot {\bf p} +p_0 \openone \right ) \phi_{_R} ({\bf p}) &=&
0\quad,\label{r1}\\
\left ({\bf J}\cdot {\bf p} - p_0 \openone \right ) \phi_{_L} ({\bf p})
&=& 0 \label{l1}
\end{eqnarray}
\end{mathletters}
have acausal dispersion relations, see Table 2 in~\cite{DVA01}. In the
case of the spin $j=1$ this manifests in existence of the solution with
the energy $E=0$.  Some time ago we learned that the same problem has been
discussed by J. R. Oppenheimer~\cite{OPPEN}, S. Weinberg~[71b]
and E. Gianetto~[49c]. For instance, Weinberg on p. 888
indicated that ``for $j=1/2$  [the equations (\ref{r1},\ref{l1})] are the
Weyl equations for the left- and right-handed neutrino fields, while for
$j=1$ they are just Maxwell's free-space equations for left- and
right-circularly polarized radiation:
\begin{mathletters}
\begin{eqnarray}
\bbox{\nabla} \times [{\bf E} -i {\bf B} ] +i
(\partial/\partial t) [{\bf E} - i{\bf B}] &=& 0,\label{mel}\\
\bbox{\nabla} \times
[{\bf E} +i {\bf B} ] -i (\partial/\partial t) [{\bf E} + i{\bf B}]  &=&
0.  \label{mer}
\end{eqnarray} \end{mathletters} The fact that these field equations
are of first order for any spin seems to me to be of no great
significance, since in the case of massive particles we can get along
perfectly well with $(2j+1)-$ component fields which satisfy only
the Klein-Gordon equation." This is obviously a remarkable and bold
conclusion of the great physicist. In the rest of the paper we try to
understand it.

Oppenheimer  concerns with the $E=0$ solution on the pages 729, 730, 733
(see also the discussion on p. 735) and indicated at its connection with
the electrostatic solutions of Maxwell's equations. ``In the absence of
charges there may be no such field." This is contradictory: free-space
Maxwell's equations do not contain terms $\rho_e$ or $\rho_m$, the charge
densities, but dispersion relations still tell us about the solution
$E=0$.  He deals further with the matters of relativistic invariance of
the matrix equation (p. 733) and suggests that the components of $\psi$
($\phi_{_{R,L}}$  in the notation of~\cite{DVA01,DVA02}) transform under a
pure Lorentz transformations like the space components of a covariant
4-vector. This induces him to extend the matrices and the wave functions
to include the forth component. Similar formulation has been developed by
Majorana~\cite{MAJOR1}. If so, it would be already difficult to consider
$\phi_{_{R,L}}$ as Helmoltz' bivectors because they have different laws
for pure Lorentz transformations.  What does the 4-component function (and
its space components) corresponds to?  Finally, he indicated (p. 728) that
$c{\bbox \tau}$, the angular momentum matrices, and the corresponding
density-flux vector may ``play in some respects the part of the velocity",
with eigenvalues $0, \pm c$. So, the formula (5) of the paper~\cite{OPPEN}
may have some relations with the discussion of the convection displacement
current in~\cite{CHUBYK2}, see below.

Finally, M. Moshinsky and A. Del Sol found the solution of the similar
nature in a two-body relativistic problem~\cite{MOSHIN}. Of course, it is
connected with earlier considerations, {\it e.g.}, in the quasipotential
approach.  In order to try to understand the physical sense of the $E=0$
solutions and the corresponding field components let us consider other
generalizations of the Maxwell's formalism.

\smallskip

{\it The formalism `baroque'.}  In this formalism proposed in
the fifties by K. Imaeda~\cite{IMAEDA} and T. Ohmura~\cite{OHMURA},
who intended to solve the problem of stability of an electron, additional
scalar and pseudoscalar fields are introduced in the Maxwell's theory.
Monopoles and magnetic currents are also present in this theory.  The
equations become:
\begin{mathletters} \begin{eqnarray}
&&\mbox{rot}\,
{\bf H} - \partial {\bf E}/\partial x_0 = {\bf i} -\mbox{grad}\, e
\quad,\label{1a}\\
&&\mbox{rot}\, {\bf E} +\partial {\bf H} /\partial x_0 = {\bf j}
+\mbox{grad}\, h\quad,\label{1b}\\
&&\mbox{div}\, {\bf E} = \rho +\partial
e/ \partial x_0\quad,\label{1c}\\ &&\mbox{div}\, {\bf H} = -\sigma
+\partial h/\partial x_0\quad.\label{1d}
\end{eqnarray} \end{mathletters}
``Each of ${\bf E}$ and ${\bf H}$  is separated into two parts ${\bf
E}^{(1)} + {\bf E}^{(2)}$ and ${\bf H}^{(1)} +{\bf H}^{(2)}$:  one is the
solution of the equations with ${\bf j},\,\sigma,\, h$ zero, and other is
the solution of the equations with ${\bf i},\, \rho,\, e$ zero."
Furthermore, Dra. T. Ohmura indicated at existence of
longitudinal photons in her model: ``It will be interesting to test
experimentally whether the $\gamma$-ray keeps on its transverse property
even in the high energy region as derived from the Maxwell theory or it
does not as predicted from our hypothesis." In fact,
the equations (\ref{1a}-\ref{1d}) can be written in a matrix notation,
what leads to the known Majorana-Oppenheimer formalism for the
$(0,0)\oplus (1,0)$ (or $(0,0)\oplus (0,1)$) representation of the
Poincar\`e group~\cite{MAJOR1,OPPEN}, see also~\cite{DOWKER}.
In a form with the Majorana-Oppenheimer matrices
\begin{mathletters}
\begin{eqnarray} \rho^1 &=&\pmatrix{0&-1&0&0\cr -1&0&0&0\cr 0&0&0&-i\cr
0&0&i&0\cr}\quad,\quad\rho^2 =\pmatrix{0&0&-1&0\cr 0&0&0&i\cr -1&0&0&0\cr
0&-i&0&0\cr}\quad,\quad \\ \rho^3 &=&\pmatrix{0&0&0&-1\cr 0&0&-i&0\cr
0&i&0&0\cr -1&0&0&0\cr}\quad, \quad\rho^0 = \openone_{4\times 4}\quad,
\end{eqnarray} \end{mathletters} and $\overline{\rho}^0 \equiv\rho^0$\,\,
,\,\, $\overline{\rho}^i \equiv -\rho^i$ the equations without an explicit
mass term are written \begin{mathletters} \begin{eqnarray} (\rho^\mu
\partial_\mu) \psi_1 (x) &=& \phi_1 (x)\quad,\\ (\overline{\rho}^\mu
\partial_\mu) \psi_2 (x) &=& \phi_2 (x)\quad.
\end{eqnarray}
\end{mathletters}
The $\phi_i$ are the ``quadrivectors" of the sources
\begin{equation}
\phi_1 = \pmatrix{-\rho +i\sigma\cr i{\bf j} - {\bf
i}\cr}\quad,\quad \phi_2 = \pmatrix{\rho +i\sigma\cr -i{\bf j} - {\bf
i}\cr}\quad.
\end{equation}
The field functions are
\begin{equation} \psi_1 (p^\mu) = {\cal C} \psi_2^\ast (p^\mu) =
\pmatrix{-i (E^0 + iB^0)\cr E^1 + iB^1\cr E^2 + iB^2\cr E^3 + iB^3\cr}\,\,
,\,\, \psi_2 (p^\mu) = {\cal C} \psi_1^\ast (p^\mu) = \pmatrix{-i (E^0 -
iB^0)\cr E^1 - iB^1\cr E^2 - iB^2\cr E^3 - iB^3\cr}\, ,
\end{equation}
where  $E^0 \equiv -h$, \, $B^0 \equiv e$ and
\begin{equation}
{\cal C} = {\cal C}^{-1} =\pmatrix{-1&0&0&0\cr 0&1&0&0\cr 0&0&1&0\cr
0&0&0&1\cr}\quad,\quad {\cal C}\alpha^\mu {\cal C}^{-1} =
\overline{\alpha}^{\mu\,\,\ast}\quad.
\end{equation}
When sources are switched off the equations have relativistic dispersion
relations $E=\pm \vert {\bf p} \vert$ only. In ref.~\cite{MAJOR1}
zero-components of $\psi$ have been connected with $\pi_0=\partial_\mu
A^\mu$, the zero-component of the canonically conjugate momentum to the
field $A_\mu$.  H. E. Moses developed the Oppenheimer's idea~\cite{OPPEN}
that the longitudinal part of the electromagnetic field is connected
somehow with the sources which created it~\cite[Eq.(5.21)]{MOSES}.
Moreover, it was mentioned in this work that even after the switchoff
of the sources, the localized field can possess the longitudinal component
({\it Example 2}). Then, he made a convention which, in my opinion, is
required to give more rigorous scientific basis:  ``\ldots $\psi^A$ is not
suitable for a final field because it is not purely transverse.  Hence we
shall subtract the part whose divergence is not zero."

Finally, one should mention ref.~\cite{BONDI}, the proposed formalism is
connected with the formalism of the  previously cited works (and with
the massive Proca theory) .  Two of Maxwell's equations remain unchanged,
but one has additional terms in two other ones:
\begin{mathletters}
\begin{eqnarray} &&{\bbox \nabla}\times {\bf H} - \partial {\bf D}/
\partial t = {\bf J} - (1/\mu_0 l^2) {\bf A}\quad,\\ &&{\bbox\nabla} \cdot
{\bf D} =\rho -(\epsilon_0/l^2) V\quad, \end{eqnarray} \end{mathletters}
where $l$ is of the dimensions length and is suggested by Lyttleton and
Bondi to be of the order of the radius of the Universe. ${\bf A}$ and $V$
are the vector and scalar potentials, which put back into two Maxwell's
equations for strengths. So, these additional terms contain information
about possible effects of the photon mass. This was applied  to explain
the expansion of the Universe.  The Watson's generalization, also
discussed in~[48b], is based on the introduction of the additional
gradient current (as in Eqs.  (\ref{1a},\ref{1c}))  and, in fact, repeats
in essence the Majorana-Oppenheimer and Imaeda-Ohmura formulations.  On a
scale much smaller than a radius of the Universe, both formulations were
shown by Chambers to be equivalent. The difference obtained is of
order $l^{-2}$ at the most. In fact, both formulations were noted by
Chambers to contain local creation of the charge.\footnote{The question of
the integral conservation of the charge over the volume still deserves
elaboration, the question of possibility to observe such a type of
non-conservation as well. These questions may be connected with the
boundary conditions on the sphere of the radius $l$.}

\smallskip

{\it  The theory of Elementary Matter.} The formalism proposed by M.
Sachs~\cite{SACHS1,SACHS2} is on the basis of the consideration of
spinorial functions composed of  3-vector components:
\begin{equation}
\phi_1 = \pmatrix{G_3\cr G_1 +iG_2\cr}\quad,\quad
\phi_2 = \pmatrix{G_1 -iG_2\cr -G_3\cr}\quad,
\end{equation}
where $G_k =H_k +iE_k$ ($k=1,2,3$).
2-component functions of the currents are constructed in the following way:
\begin{equation}
\Upsilon_1 =-4\pi i \pmatrix{\rho +j_3\cr j_1 + ij_2 \cr}\quad,\quad
\Upsilon_2 = -4\pi i \pmatrix{j_1 - ij_2 \cr \rho -j_3 \cr}\quad.
\end{equation}
The dynamical equation in this formalism reads
\begin{equation}
\sigma^\mu \partial_\mu \phi_\alpha = \Upsilon_\alpha\quad.\label{seq}
\end{equation}
``\ldots Eq. (\ref{seq}) is {\it not} equivalent to the less general
form of Maxwell's equations. That is to say the spinor equations
(\ref{seq}) are not merely a rewriting of the vector form of the field
equations, they are a true generalization in the sense of transcending the
predictions of the older form while also agreeing with all of the correct
predictions of the latter \ldots Eq. (\ref{seq}) may be rewritten
in the form of four conservation equations
$\partial_\mu (\phi_\alpha^\dagger \sigma^\mu \phi_\beta ) =
\phi_\alpha^\dagger \Upsilon_\beta +\Upsilon_\alpha^\dagger \phi_\beta$
[which] entails eight real conservation laws." For instance, these
equations could serve as a basis for describing parity-violating
interactions~[60a], and can account for the spin-spin
interaction as well from the beginning~[60d,p.934]. The formalism
was applied to explain several puzzles in neutrino physics. The
connection with the Pauli Exclusion Principle was revealed.  The theory,
when the interaction (`matter field labeling') is included, is essentially
bi-local.\footnote{The hypothesis of the non-local nature of the charge
seems to have been firstly proposed by J.~Frenkel.} ``What was discovered
in this research program, applied to the particle-antiparticle pair, was
that an exact solution for the coupled field equations for the pair, in
its rest frame, gives rise (from Noether's theorem) to a prediction of null
energy, momentum and angular momentum, when it is in this particular {\it
bound state}~\cite{SACHS2}." Later~\cite{SACHS2} this type of equations
was written in the quaternion form with the continuous function $m=\lambda
\hbar /c$ identified with the inertial mass. Thus, an extension of the
model to the general relativity case was proposed.  Physical consequences
of the theory are: a) the formalism predicts while small but non-zero
masses and the infinite  spectrum of neutrinos; b) the Planck spectral
distribution of black body radiation follows; c) the hydrogen spectrum
(including the Lamb shift) can be deduced; d) grounds for the charge
quantization are proposed; e) the lifetime  of the muon state was
predicted; f) the electron-muon mass splitting was discussed, ``the
difference in the mass eigenvalues of a doublet depends on the alteration
of the geometry of space-time in the vicinity of excited pairs of the
``physical vacuum" [``a degenerate gas of spin-zero objects", longitudinal
and scalar photons, in fact -- my comment] --- leading, in turn, to a
dependence of the ratio of mass eigenvalues on the fine-structure
constant". That was impressive work and these are impressive results!

\smallskip

{\it Quantum mechanics of the phase.} A.
Staruszkiewicz~\cite{STARU1,STARU2} considers the Lagrangian and the
action of a potential formulation for the electromagnetic field, which
include a longitudinal part:
\begin{equation}
{\cal S} = -{1\over 16\pi}
\int d^4 x \left \{ F_{\mu\nu} F^{\mu\nu} +2\gamma \left ( \partial^\mu
A_\mu +{1\over e} {\,\lower0.9pt\vbox{\hrule \hbox{\vrule height 0.2 cm
\hskip 0.2 cm \vrule height 0.2cm}\hrule}\,} S \right )^2 \right \} \quad.
\end{equation}
$S$ is a scalar field called the phase. As a matter  of fact, this
formulation was shown to be a development of the Dirac-Fock-Podol'sky
model in which the current is a gradient of some scalar field~\cite{FOCK}:
\begin{equation}
4\pi j_\nu = - \partial_\nu F\quad.
\end{equation}
The modified Maxwell's equations are written:
\begin{mathletters}
\begin{eqnarray}
&&\partial_\lambda F_{\mu\nu} +\partial_\mu F_{\nu\lambda} +\partial_\nu
F_{\lambda\mu} =0\quad,\\
&& \partial^\mu F_{\mu\nu} +\partial_\nu F =0\quad.
\end{eqnarray} \end{mathletters}
Again we see a gradient current and, therefore, the Dirac-Fock-Podol'sky
model is a simplified version (seems, without monopoles) of the more
general Majorana-Oppenheimer theory. Staruszkiewicz put forth the
questions~\cite{STARU2}, see also~[57b] and ~\cite{GERSTEN}:
``Is it possible to have a system, whose motion is determined completely
by the charge conservation law alone?  Is it possible to have a pure
charge not attached to a nonelectromagnetic piece of matter?" and
answering came to the conclusion ``that the Maxwell electrodynamics of a
gradient current is a closed dynamical system." The interpretation of a
scalar field as a phase of the expansion motion of a charge under
repulsive electromagnetic forces was proposed.  ``They [the
Dirac-Fock-Podol'sky  equations] describe a charge let loose by removal of
the Poincar\`e stresses." The phase was then related with the vector
potential by means of~[64e,p.902]\,\footnote{The formula (\ref{phas}) is
reminiscent to the Barut self-field electrodynamics~\cite{BARUT1}. This
should be investigated by taking 4-divergence of the Barut's {\it
anzatz}.} \begin{equation} S(x) = -e \int A_\mu (x -y) j^\mu (y) d^4
y\quad,\quad \partial_\mu j^\mu (y) = \delta^{(4)} (y)\quad. \label{phas}
\end{equation} The operator of a number of zero-frequency photons was
studied. The total charge of the system, found on the basis of the
N\"other theorem, was connected with the change of the phase between the
positive and the negative time-like infinity:  $Q =-{e\over 4\pi} \left [
S(+\infty) -S(-\infty) \right ]$. It was shown that $e^{iS}$, having a
Bose-Einstein statistics, can serve itself as a creation operator: $Q
e^{iS} \vert 0> = \left [ Q, e^{iS} \right ]\vert 0> = -e \,\,e^{iS} \vert
0 >$.  Questions of fixing  the factor $\gamma$ by appropriate physical
conditions were also answered. Finally, the Coulomb field was decomposed
into irreducible unitary representations of the proper orthochronous
Lorentz group~\cite{STARU3}. Both representations of the main serie and
the supplementary serie were regarded.  In my opinion, these researches
can help to understand the nature of the charge and of the fine structure
constant.

\smallskip

{\it Invariant evolution parameter.} The theory of electromagnetic field
with an invariant evolution parameter ($\tau$ , the Newtonian time) has
been worked out by L. P.  Horwitz~\cite{HORWITZ1,HORWITZ2,HORWITZ3}.
It is a development of the Stueckelberg formalism~\cite{STUCKEL}
and I consider this theory as an important step to understanding the
nature of our space-time. The Stueckelberg  equation:
\begin{equation}
i{\partial \psi_\tau (x) \over \partial \tau} = K \psi_\tau (x)\label{ss}
\end{equation}
is deduced on the basis of his worldline classical relativistic mechanics
with following setting up the covariant commutation relations $[x^\mu,
\, p^\nu] = ig^{\mu\nu}$.  Remarkably that he proposed a classical
analogue of antiparticle (which, in fact, has been later used by R.
Feynman) and of annihilation processes. As noted by Horwitz if one insists
on the $U(1)$ gauge invariance of the theory based on the
Stueckelberg-Schr\"odinger equation (\ref{ss}) one arrives at
the 5-potential electrodynamics ($i\partial_\tau \rightarrow
i\partial_\tau +e_0 a_5$) where the equations, which are deduced by means
of the variational principle, read \begin{equation} \partial_\beta
f^{\alpha\beta} = j^\alpha \label{hor} \end{equation} ($\alpha,\beta =
1\ldots 5$), with an additional fifth component of the conserved current
$\rho = \vert \psi_\tau (x)\vert^2$.  The underlying symmetry of the
theory can be $O(3,2)$ or $O(4,1)$ ``depending on the choice of metric for
the raising and lowering of the fifth ($\tau$) index~\cite{HORWITZ1}".
For Minkowski-space components the equation (\ref{hor}) is reduced to
$\partial_\nu f^{\mu\nu} +\partial_\tau f^{\mu 5} = j^\mu$. The Maxwell's
theory is recovered after integrating over $\tau$ from $-\infty$ to
$\infty$, with appropriate asymptotic conditions.  The formalism has been
applied mainly in the study of the many-body problem and in the
measurement theory, namely, bound states (the hydrogen atom), the
scattering problem, the calculation of the selection rules and amplitudes
for radiative decays, a covariant Zeeman effect, the Landau-Peierls
inequality.  Two crucial experiments which may check validity and
may distinguish the theory from ordinary approaches  have also been
proposed~\cite[p.15]{HORWITZ3}.

Furthermore, one should  mention that in the
framework of the special relativity version of the Feynman-Dyson proof of
the Maxwell's equations~\cite{DYSON} S. Tanimura came to  unexpected
conclusions~\cite{TANIMU} which are related with the formulation defended
by L.  Horwitz.  Trying to prove the Maxwell's formalism
S. Tanimura arrived at the conclusion about a theoretical
possibility of its generalization.  According to his consideration the
4-force acting on the particle in the electromagnetic field must be
expressed in terms of
\begin{equation} F^\mu (x,\dot x) = G^\mu (x) +
< F^\mu_{\quad \nu} (x) \,\,\dot x^\nu >\quad,
\end{equation}
where the symbol
$<\ldots >$ refers to the Weyl-ordering prescription.  The fields $G^\mu
(x)$ , $F^\mu_{\quad\nu} (x)$ satisfy\footnote{Of course, one can repeat
the Tanimura proof for dual fields and obtain two additional equations.}
\begin{mathletters}
\begin{eqnarray}
&&\partial_\mu G_\nu -\partial_\nu G_\mu =0\quad,\\
&&\partial_\mu F_{\nu\rho} +\partial_\nu F_{\rho\mu} +\partial_\rho
F_{\mu\nu} =0 \quad.
\end{eqnarray} \end{mathletters}
This implies that apart
from the 4-vector potential $F_{\mu\nu} =\partial_\mu A_\nu -\partial_\nu
A_\mu$ there exists a scalar field $\phi (x)$ such that $G_\mu
=\partial_\mu \phi$.  One may try to compare this result with the fact of
existence of additional scalar field
components in the Majorana-Oppenheimer formulation of
electrodynamics and with the Stueckelberg-Horwitz theory.  The latter has
been done by Prof.  Horwitz himself~[38c] by the
identification $F_{\mu 5} = - F_{5\mu} = G_\mu$ and the explicit
demonstration that for the off-shell theory the Tanimura's equations
reduce to
\begin{mathletters} \begin{eqnarray} &&\partial_\mu F_{\nu\rho}
+\partial_\nu F_{\rho\sigma} +\partial_\rho F_{\mu\nu} =0 \quad,\\
&&\partial_\mu G_\nu -\partial_\nu G_\mu +{\partial F_{\mu\nu} \over
\partial \tau} =0\quad,\\ && m\ddot x^\mu = G^\mu (\tau, x) + F^{\mu\nu}
(\tau, x) \dot x_\nu\quad.  \end{eqnarray} \end{mathletters}

Finally, among theories with additional parameters one should mention the
quantum field model built in the de Sitter momentum space
$p_5^2 -p_4^2 - p_3^2 -p_2^2 -p_1^2 = M^2$, ref.~\cite{KADYSH}. The
parameter $M$ is considered as a new physical constant,
the fundamental mass. In a configurational space  defined on the basis of
the Shapiro transfromations the equations become the finite-difference
equations thus leading to the lattice
structure of the space. In the low-energy limit ($M\rightarrow \infty$)
the theory is equivalent to the standard one.

\smallskip

{\it Action-at-a-distance.} In the paper~\cite{CHUBYK1} A. E. Chubykalo and
R. Smirnov-Rueda revealed on the basis of the analysis of the Cauchy
problem of the D'Alembert and the Poisson equations  that one should
revive the concept of the {\it instantaneous} action-at-a-distance in
classical electrodynamics in order to remove some inconsistencies of the
description by means of the Li\'enard-Wiechert potentials. The essential
feature of the formalism is in introduction of two types of field
functions, with the explicit and implicit dependences on time. The
energy of longitudinal modes in this formulation cannot be stored locally
in the space, the spread velocity may be whatever and so one has also
$E=0$.  The new convection displacement current was proposed
in~\cite{CHUBYK2} on the basis of the development of this wisdom.  It has
a form $j_{\mbox{\small{disp}}} = -{1\over 4\pi} ({\bf v} \cdot
\bbox{\nabla} ) {\bf E}$.  This is a resurrection of the Hertz' ideas
(later these ideas have been defended by T.  E. Phipps, jr.) to replace
the partial derivative by the total derivative in the Maxwell's equations.
In my opinion, one can also reveal connections with the
Majorana-Oppenheimer formulation following to the analysis of
ref.~\cite[p.728]{OPPEN}.

F. Belinfante~[11a] appears to come even earlier to the Sachs' idea
about the ``physical vacuum" as pairs of some particles from a
very different viewpoint. In his formulation of
the quantum-electrodynamic perturbation theory
zero-order approximation is determined
in which scalar and longitudinal photons are present in pairs.  He also
considered~[11b] the Coulomb problem in the frameworks of the quantum
electrodynamics and proved that the signal can be transmitted with the
velocity greater than $c$. So, this old work appears to be in accordance
with recent experimetal data (particularly, with the claims of G. Nimtz
{\it et al.} about a wave packet propagating faster than $c$ through a
barrier, which was used ``to transmit Mozart's Symphony No. 40 through a
tunnel of $114 \,mm$ length at a speed of $4.7c$").  As indicated by E.
Recami in a private communication the $E=0$ solutions can be put in
correspondence to a tachyon of the infinite velocity.

\smallskip

{\it Evans-Vigier ${\bf B}^{(3)}$ field.} In a recent serie of remarkable
papers (in FPL, FP, Physica A and B, Nuovo Cimento B) and books M.
Evans and J.-P.  Vigier indicated the possibility of consideration of
the longitudinal ${\bf B}^{(3)}$ field for describing many electromagnetic
phenomena and in cosmological models as well~\cite{EVANS1}. It is
connected with transversal modes
\begin{mathletters} \begin{eqnarray} {\bf
B}^{(1)} &=&\frac{B^{(0)}}{\sqrt{2}} (i\,{\bf i} + {\bf j})
\,e^{i\phi}\quad,\\ {\bf B}^{(2)} &=&\frac{B^{(0)}}{\sqrt{2}} (-i\,{\bf i}
+ {\bf j}) \,e^{-i\phi}\quad, \end{eqnarray} \end{mathletters} $\phi
=\omega t -{\bf k}\cdot {\bf r}$, by means of the ciclic relations
\begin{mathletters} \begin{eqnarray} {\bf B}^{(1)} \times {\bf B}^{(2)} =
\, i B^{(0)} {\bf B}^{(3)\,*}\quad,\label{e1}\\ {\bf B}^{(2)} \times {\bf
B}^{(3)} = \, i B^{(0)} {\bf B}^{(1)\,*}\quad,\label{e2}\\ {\bf B}^{(3)}
\times {\bf B}^{(1)} = \, i B^{(0)} {\bf B}^{(2)\,*}\quad.  \label{e3}
\end{eqnarray} \end{mathletters} The indices $(1),(2),(3)$ denote
components of the vector in the circular basis and, thus, the longitudinal
field ${\bf B}^{(3)}$ presents itself a third component of the  3-vector
in some isovector space. ``The conventional $O(2)$ gauge geometry is
replaced by a non-Abelian $O(3)$ gauge geometry and the Maxwell equations
are thereby generalized" in this approach. Furthermore, some success in
the problem of the unification of gravitation and electromagnetism has
been achieved in recent papers by M. Evans. It was pointed out by
several authors,  {\it e.g.},~\cite{EVANS2,DVO5} that this field is the
simplest and most natural (classical) representation of a particles spin,
the additional phase-free discrete variable discussed by
Wigner~\cite{WIGNER}.  The consideration by Y.~S.  Kim {\it et al.}, see
ref.~[36a,formula (14)], ensures that the problem of physical significance
of Evans-Vigier-type longitudinal modes is related with the problem of the
normalization and of existence of the mass of a particle transformed on
the $(1,0)\oplus (0,1)$ representation of the Poincar\`e group.
Considering explicit forms of the $(1,0)\oplus (0,1)$ ``bispinors" in the
light-front formulation~\cite{DIRAC1} of the quantum field theory of this
representation (the Weinberg-Soper formalism) D. V.  Ahluwalia and M.
Sawicki showed that in the massless limit one has only two non-vanishing
Dirac-like solutions. The ``bispinor" corresponding to the longitudinal
solution is directly proportional to the mass of the particle.  So, the
massless limit of this theory, the relevance of the $E(2)$ group to
describing physical phenomena and the problem of what is mass deserve
further consideration.

\smallskip

{\it Antisymmetric tensor fields.} To my knowledge  researches of
antisymmetric tensor fields in the quantum theory began from the paper by
V. I. Ogievetski\u{\i} and I. V. Polubarinov~\cite{OGIEVET}.  They
claimed that the antisymmetric tensor field ({\it notoph} in the
terminology used, which I find quite suitable) can
be longitudinal in the quantum theory, owing to the new gauge invariance
\begin{equation}
F_{\mu\nu} \rightarrow F_{\mu\nu} +\partial_\mu \Lambda
-\partial_\nu \Lambda \quad
\end{equation}
and applications of the supplementary conditions. The result by
Ogievetski\u{\i} and Polubarinov has been repeated by
K. Hayashi~\cite{HAYASHI}, M. Kalb and P. Ramond~\cite{KALB} and T. E.
Clark  {\it et al.}~\cite{LOVE}.   The Lagrangian
($F_k=i\epsilon_{kjmn}F_{jm,n}$)
\begin{equation} {\cal L}^{H}=\frac{1}{8}
F_k F_k = -{1\over 4}(\partial_{\mu} F_{\nu\alpha}) (\partial_{\mu}
F_{\nu\alpha})+ {1\over 2} (\partial_{\mu} F_{\nu\alpha})(\partial_{\nu}
F_{\mu\alpha}) \end{equation}
after the application of the Fermi method
{\it mutatis mutandis} (comparing with the case of
the 4-vector potential field) yields the spin
dynamical invariant to be equal to zero.  While several authors insisted
on the transversality of the antisymmetric tensor field and the necessity
of gauge-independent consideration~\cite{GURSEY,TAKAHA,BOYAR} perpetually
this interpretation (`longitudity') has become wide-accepted.
In refs.~\cite{AVD1,AVD2} an antisymmetric tensor {\it
matter} field was studied and it appears to be also longitudinal, but
to have two degrees of freedom.   Unfortunately, the authors of the
cited work regarded only a massless {\it real} field and did not take into
account the physical reality of the dual field corresponding to an
antiparticle.  But, what is important,  L. Avdeev and M. Chizhov
noted~\cite{AVD2} that in such a framework there exist $\delta^\prime$
transversal solutions, which cannot be interpreted as relativistic
particles.

If an  antisymmetric tensor field would be pure longitudinal, it appears
failure to understand, why in the classical electromagnetism we
are convinced that an antisymmetric tensor field is transversal. Does this
signifies one must abandon the Correspondence Principle? Moreover, this
result contradicts with the Weinberg theorem $B-A=\lambda$, ref.~[71b].
This situation has been later analyzed in
refs.~\cite{DVOOLD,EVANS2,DVO2,DVO5,DVO6} and it was found that indeed the
``longitudinal nature" of antisymmetric tensor fields is connected with
the application of the generalized Lorentz condition to the quantum
states: $\partial_\mu F^{\mu\nu} \vert \Psi > =0$.  Such a procedure leads
also (like in the case of the treatment of the 4-vector potential
field without proper regarding the phase field) to the problem of the
indefinite metric which was noted by Gupta and Bleuler.
So, it is already obviously from methodological
viewpoints that the grounds for regarding only particular cases can be
doubted by  the Lorentz symmetry principles.  Ignoring the phase field of
Dirac-Fock-Podol'sky-Staruszkiewicz or ignoring $\chi$
functions~\cite{DVO4} related with the 4-current and, hence, with
the non-zero value of $\partial_\mu F^{\mu\nu}$ can put obstacles in the
way of creation of the unified field theory and embarass understanding the
physical content dictated by the Relativity Theory.

\section{The Weinberg formalism}

In the beginning of the sixties the $2(2j+1)$- component approach
has been proposed in order to construct a Lorentz-invariant
interaction $S$-matrix  from the first
principles~\cite{BARUT0,JOOS,WEIN1,WEAVER,WEIN2,MARINOV,HAMMER}.  The
authors  had thus some hopes on an adequate perturbation calculus for
processes including higher-spin particles which appeared in disposition of
physicists in that time. The field theory in that time was in  some
troubles.

The Weinberg {\it anzatzen} for the $(j,0)\oplus (0,j)$ field theory
are simple and obvious~[71a,p.B1318]:\\
a) relativistic invariance
\begin{equation} U [\Lambda, a] \psi_n (x) U^{-1} [\Lambda, a]
= \sum_m D_{nm} [\Lambda^{-1}] \psi_m (\Lambda x +a)\quad, \label{ri}
\end{equation}
where $D_{nm} [\Lambda]$ is the corresponding representation of
$\Lambda$;\\ b) causality
\begin{equation}
\left [ \psi_n (x), \psi_m (y) \right ]_{\pm} = 0\quad
\end{equation}
for $(x-y)$ spacelike, which garantees the commutator
of the Hamiltonian density $[{\cal H} (x), {\cal H} (y)] =0$, provided
that ${\cal H} (x)$ contains an even number of fermion field factors.
The interaction Hamiltonian ${\cal H} (x)$ is constructed out of the
creation and annihilation operators for the free particles described by
some $H_0$, the free-particle part of the Hamiltonian.  Thus, the
$(j,0)\oplus (0,j)$ field \begin{equation} \psi (x) = \pmatrix{\varphi
(x)\cr \chi (x)\cr} \end{equation} transforms according to (\ref{ri}),
where \begin{equation} {\cal D}^{(j)} [ \Lambda ] = \pmatrix{D^{(j)} [
\Lambda ]&0\cr 0& \overline D^{(j)} [\Lambda ]\cr},\quad D^{(j)} [\Lambda
] = \overline D^{(j)} [\Lambda^{-1}]^\dagger \quad,\quad {\cal D}^{(j)}
[\Lambda ]^\dagger = \beta {\cal D}^{(j)} [\Lambda^{-1} ] \beta \,\,
,\label{lt} \end{equation} with \begin{equation} \beta
=\pmatrix{0&\openone\cr \openone &0\cr}\quad, \end{equation}
and, hence, for
pure Lorentz transformations (boosts) \begin{mathletters} \begin{eqnarray}
D^{(j)} [L ({\bf p}) ] &=& \exp (-\hat p\cdot {\bf J}^{(j)} \theta)\quad,
\label{lt1}\\
\overline D^{(j)} [L ({\bf p}) ] &=& \exp (+\hat p\cdot {\bf J}^{(j)}
\theta)\quad,\label{lt2}
\end{eqnarray} \end{mathletters}
with $\sinh \theta \equiv \vert {\bf p}\vert /m$.
Dynamical equations, which Weinberg proposed, are (Eqs. (7.17) and (7.18)
of the first paper~\cite{WEIN1}):
\begin{mathletters} \begin{eqnarray}
\overline \Pi (-i\partial) \varphi (x) &=& m^{2j} \chi (x)\quad,\\
\Pi (-i\partial) \chi (x) &=& m^{2j} \varphi (x)\quad.
\end{eqnarray}
\end{mathletters}
They are rewritten into the form  (Eq. (7.19) of~[71a])
\begin{equation}
\left [\gamma^{\mu_1 \mu_2 \ldots \mu_{2j}} \partial_{\mu_1}
\partial_{\mu_2} \ldots \partial_{\mu_{2j}} +m^{2j} \right ] \psi (x) =0
\quad,
\end{equation}
with the Barut-Muzinich-Williams matrices ~\cite{BARUT0}
\begin{equation}
\gamma^{\mu_1 \mu_2 \ldots \mu_{2j}} = -i^{2j} \pmatrix{0&t^{\mu_1 \mu_2
\ldots \mu_{2j}}\cr
\overline{t}^{\mu_1 \mu_2 \ldots \mu_{2j}} &0\cr}\quad.
\end{equation}
The following notation was used
\begin{equation}
\Pi_{\sigma^\prime \sigma}^{(j)} (q) \equiv (-1)^{2j} t_{\sigma^\prime
\sigma}^{\quad \mu_1 \mu_2 \ldots \mu_{2j}} q_{\mu_1} q_{\mu_2} \ldots
q_{\mu_{2j}}\quad,\quad
\end{equation}
\begin{equation}
\overline \Pi^{(j)}_{\sigma^\prime \sigma} (q) \equiv (-1)^{2j} \overline
t^{\mu_1 \mu_2 \ldots \mu_{2j}}_{\sigma^\prime \sigma} q_{\mu_1} q_{\mu_2}
\ldots q_{\mu_{2j}}\quad,\quad \overline \Pi^{(j)\,\ast} (q)  =C \Pi^{(j)}
C^{-1}\quad,
\end{equation}
with $C$ being the matrix of the charge conjugation in the $2j+1$-
dimension representation.  The tensor $t$ is defined in a following manner:

\begin{itemize}
\item
$t_{\sigma^\prime \sigma}^{\quad \mu_1 \mu_2 \ldots \mu_{2j}}$ is a
$2j+1$  matrix with $\sigma , \sigma^\prime =j, j-1,\ldots -j$; $\mu_1 ,
\mu_2 \ldots \mu_{2j} =0, 1, 2, 3$;

\item
$t$ is symmetric in all $\mu$'s;

\item
$t$ is traceless in all $\mu$'s, {\it i.e.}, $g_{\mu_1 \mu_2}
t_{\sigma^\prime \sigma}^{\quad \mu_1 \mu_2 \ldots \mu_{2j}}$ and with all
permutations of upper indices;

\item
$t$ is a tensor under Lorentz transformations,
\begin{mathletters}
\begin{eqnarray}
D^{(j)}[\Lambda] t^{\ \mu_1 \mu_2 \ldots \mu_{2j}}
D^{(j)}[\Lambda]^{\dagger}&=&\Lambda_{\nu_1}^{\,\,\,\mu_1}
\Lambda_{\nu_2}^{\,\,\,\mu_2}\ldots
\Lambda_{\nu_{2j}}^{\,\,\,\mu_{2j}} t^{\ \nu_1 \nu_2 \ldots
\nu_{2j}}\quad,\\
\overline D^{(j)}[\Lambda] \bar t^{\ \mu_1 \mu_2
\ldots \mu_{2j}} \overline
D^{(j)}[\Lambda]^{\dagger}&=&\Lambda_{\nu_1}^{\,\,\,\mu_1}
\Lambda_{\nu_2}^{\,\,\,\mu_2}\ldots
\Lambda_{\nu_{2j}}^{\,\,\,\mu_{2j}}\overline t^{\ \nu_1 \nu_2 \ldots
\nu_{2S}}\quad.
\end{eqnarray} \end{mathletters}
For instance, in the $j=1$ case $t^{00} =\openone$,\, $t^{0i}=t^{i0}=J_i$
and $t^{ij} = \left \{ J_i, J_j\right \} -\delta_{ij}$, with $J_i$ being
the $j=1$ spin matrices and the metric $g_{\mu\nu} = \mbox{diag}
(-1,1,1,1)$ being used. Furthermore, for this representation
\begin{equation} \overline t^{\mu_1
\mu_2 \ldots \mu_{2j}} = \pm t^{\mu_1 \mu_2 \ldots \mu_{2j}}\quad,
\end{equation} the
sign being $+1$ or $-1$ according to whether the $\mu$'s contain
altogether an even or an odd number of space-like indices.
\end{itemize}

The Feynman diagram technique has been built and some properties with
respect to discrete symmetry operations have been studied.
The propagator used in the Feynman diagram technique is found not to be the
propagator arising from the Wick theorem because of extra terms
proportional to equal-time $\delta$ functions and their derivatives
appearing if one uses the time-ordering product of field operators
$<T \left \{ \psi_\alpha (x) \overline \psi_\beta (y) \right \} >_0$.
The covariant propagator is defined by
\begin{eqnarray}
\lefteqn{S_{\alpha\beta} (x-y) = (2\pi)^{-3} m^{-2j} M_{\alpha\beta}
(-i\partial ) \int {d^3 {\bf p} \over 2\omega ({\bf p})}
\left [ \theta (x-y) \exp \{ip\cdot (x-y) \} + \right.\nonumber}\\
&+&\left.\theta
(y-x) \exp \{ ip\cdot (y-x) \} \right ] =-im^{-2j} M_{\alpha\beta}
(-i\partial ) \Delta^C (x-y)\quad,
\end{eqnarray}
where
\begin{equation}
M(p) = \pmatrix{m^{2j} & \Pi (p)\cr
\overline \Pi (p) & m^{2j}\cr}\quad,
\end{equation}
and $\Delta^C (x)$ is  the covariant $j=0$ propagator.

For massless particles the Weinberg theorem
about connections between the helicity of a particle and
the representation of the group $(A,B)$ which the corresponding field
transforms on has been proved. It says:  ``A massless particle operator
$a ({\bf p}, \lambda )$ of helicity $\lambda$ can only be used to
construct fields which transform according to  representations $(A, B)$,
such that $B-A=\lambda$.  For instance, a left-circularly polarized photon
with $\lambda=-1$ can be associated with $(1,0)$,  $({3\over 2}, {1\over
2})$, $(2,1)$ \ldots fields, but {\it not} with the vector potential,
$({1\over 2},{1\over 2})$\ldots  [It is not the case of a massive
particle.] A field can be constructed out of $2j+1$ operators $a({\bf p},
\sigma)$ for any representation $(A, B)$ that ``contains" $j$, such that
$j=A+B\, \mbox{or} \, A+B-1\ldots \,\mbox{or}\,\mid A-B\mid$, [{\it e.
g.}, a $j=1$ particle ] field could be a four-vector $({1\over 2},
{1\over 2})$\ldots [{\it i.e.}, built out of the vector potential ]."

In subsequent papers Weinberg showed that it is possible to construct
fields transformed on other representations of the Lorentz group but
unlikely can be considered as fundamental ones.  The prescription
for constructing fields have been given in ref.~[71c,p.1895].  ``Any
irreducible field $\psi^{(A,B)}$ for a particle of  spin $j$ may be
constructed by applying a suitable differential operator of order $2B$ to
the field $\psi^{(j,0)}$, provided that $A$, $B$, and $j$ satisfy the
triangle inequality $\vert A- B \vert \leq j \leq A+B$." For example, from
the self-dual antisymmetric tensor $F^{\mu\nu}$ the $(1/2,1/2)$ field
$\partial_\mu F^{\mu\nu}$ , the $(0,1)$ field
$\epsilon_{\mu\nu\lambda\rho}\partial^\lambda \partial_\sigma F^{\rho
\sigma}$ have been constructed.  Moreover, various invariant-type
interactions have been tabulated~[71b,p.B890] and~[71c,Section III]. While
one can also use fields from different representations of the Lorentz
group to obtain some physical predictions, in my opinion, such a wisdom
could lead us to certain mathematical inconsistencies (like the
indefinite metric problem and the subtraction of infinities~\cite{DIRAC}).
The applicability of the procedure mentioned above to massless states
should still be analyzed in detail.

Finally, we would like to cite a few paragraphs from other Weinberg's
works. In the paper of 1965 Weinberg proposed his concept
how to deal with several puzzles noted before~[72c]: ``\ldots
Tensor fields cannot by themselves be used to construct the interaction
$H^\prime (t)$, because the coefficients of the operators for creation or
annihilation of particles of momentum $p$ and spin $j$ would vanish as
$p^j$ for $p\rightarrow 0$, in contradiction with the known existence of
inverse-square-law forces. We are therefore forced to turn from these
tensor fields to the potentials\ldots The potentials are not tensor
fields; indeed, they cannot be, for we know from a very general
theorem~[71b] that no symmetric tensor field of rank $j$ can
be constructed from the creation and annihilation operators of massless
particles of spin $j$. It is for this reason that some field theorists
have been led to introduce fictitious photons and gravitons of helicity
other than $\pm j$, as well as the indefinite metric that must accompany
them. Preferring to avoid such unphysical monstrosities, we must ask
now what sort of coupling we can give our nontensor potentials without
losing the Lorentz invariance of the $S$ matrix?\ldots Those in which the
potential is coupled to a conserved current." Thus, gauge models obtain
some physical grounds from the Lorentz invariance.

\section{The Weinberg Formalism in New Development}

In the papers~\cite{SANKAR} another equation in the $(1,0)\oplus (0,1)$
representation  has been proposed. It reads  ($p_\mu$ is the differential
operator, $E=\sqrt{{\bf p}^{\,2} + m^2}$)
\begin{equation} \left [ \gamma_{\mu\nu} p_\mu p_\nu +{i(\partial
/\partial t) \over E} m^2 \right ] \psi = 0\quad.  \end{equation} The
auxiliary condition \begin{equation} (p_\mu p_\mu +m^2 ) \psi =0
\end{equation}
is implied.
``\ldots In the momentum representation the wave equation may be written
as [62b,formula (12)]
\begin{equation}
[ \pm \gamma_{\mu\nu} p_\mu p_\nu +m^2 ] U_\pm ({\bf p}) = 0\quad.
\end{equation}
corresponding to the particle and antiparticle with the column vector $U_+
({\bf p})$ and $U_- ({\bf p})$, respectively." Many dynamical
features of this approach have been analyzed in those papers but,
unfortunately, the author
erroneously claimed that the two formulation (the Weinberg's one and his
own formulation) ``are equivalent in physical content".
The matters related with the discrete symmetry operations have been
analyzed in detail in the recent years only by Ahluwalia {\it et al.},
ref.~\cite{DVA1,DVA3} and~[4b].  In this
Section first of all let us follow the arguments of the cited papers.
According to the Wigner rules (\ref{lt},\ref{lt1},\ref{lt2}) in the
notation of papers~\cite{DVA1} one has
\begin{mathletters}
\begin{eqnarray} \phi_{_R} ({\bf p}) &=& \Lambda_{_R} \phi_{_R} ({\bf 0})
= \exp (+{\bf J}\cdot\bbox{\varphi}) \phi_{_R} ({\bf 0})
\quad,\label{wig1}\\ \phi_{_L} ({\bf p}) &=& \Lambda_{_L} \phi_{_L} ({\bf
0}) = \exp (-{\bf J}\cdot\bbox{\varphi}) \phi_{_L} ({\bf 0})
\label{wig2}\quad, \end{eqnarray} \end{mathletters}
with $\phi_{_{R,L}}
({\bf p})$ being $(j,0)$ right- and $(0,j)$ left- `bispinors' in the
momentum representation, respectively; ${\bbox \varphi}$ are the
parameters of the Lorentz boost. By means of the explicit application
(\ref{wig1},\ref{wig2}) the  $u_\sigma ({\bf p})$ and $v_\sigma ({\bf p})$
bispinors have been found in the $j=1$, $j=3/2$ and $j=2$ cases~[3a]
in the generalized canonical representation. For the $(1,0)\oplus (0,1)$
`bispinors' see, {\it e.g.}, formulas (7) of~[3b].  It was proved that
the states answering for positive- and negative- energy solutions have
intrinsic parities $+1$ and $-1$, respectively, when applying the space
inversion operation. The conclusion has been achieved in the Fock
secondary-quantization space too, thus proving that we have an explicit
example of the theory envisaged by Bargmann, Wightman and Wigner (BWW)
long ago, ref.~[73b].  The remarkable feature of this formulation is: a
boson and its antiboson have opposite intrinsic parities.  Origins of
this fact have been explained in ref.~\cite{DVA02} thanks to an anonymous
referee.  Namely, ``the relative phase $\epsilon$ between particle and
antiparticle states is not arbitrary and is naturally defined as:
\begin{equation} U(P) U(C) = \epsilon U(C) U(P)\quad.  \end{equation}
[ One
can prove that ] $\epsilon =\pm 1$ by using associativity of the group
law."  Depending on the operations of the space inversion and of the
charge conjugation either commute or anticommute we  obtain
either the same or the opposite values of parities for particle and
antiparticle. The commutator of these operations in the Fock space is ``a
function of the Charge operator with formal and phenomenological
consequences", refs.~\cite{NIGAM,DVA3,DVA02}.  The massless limit of the
theory in the $(1,0)\oplus (0,1)$ representation was studied
in~\cite{DVA01,DVA02,DVA2} with the following result achieved (cited
from~\cite{DVA02}): ``Present theoretical arguments suggest that in strong
fields, or high-frequency phenomenon, Maxwell equations may not be an
adequate description of nature. Whether this is so can only be decided by
experiment(s). Similar conclusions, in apparently very different
framework, have been independently arrived at by M.  Evans and communicated
to the author."

In ref.~\cite{DVA2} the properties of the light-front-form
$(1/2,0)\oplus (0,1/2)$ bispinors have been under study.  An unexpected
result has been obtained that they do not get interchanged under the
operation of parity.  Thus, one must take into account the evolution of a
physical system not only along $x^+$ but also along the $x^-$ direction.
In~\cite{DVA3} the Majorana-McLennan-Case construct has been analized and
interesting mathematical and phenomenological connections have been found.
The analysis resulted in a serie of papers of both mine and others in many
physical journals, but a detailed presentation of these ideas is out of a
subject of this paper.\footnote{I still present references of my
work related with these problems. Here they are: V. V. Dvoeglazov, Nuovo
Cim.  A108 (1995) 1467; ibid 111B (1996) 483; Int. J. Theor. Phys. 34
(1995) 2467; ibid 35 (1996) 115; Rev. Mex. Fis. Suppl. 41 (1995) 159;
Hadronic J.  Suppl. 10 (1995) 349; ibid 10 (1995) 359; Bol. Soc. Mex. Fis.
9-3 (1995) 28; Proc. ICTE'95, Cuautitl\'an, Sept. 1995; Proc. IV Wigner
Symp., Guadalajara, Aug.  1995; Inv. Cient\'{\i}fica, in press. Several
papers are now under review in various journals.}

In a recent serie of
my papers~\cite{DVO1,DVO2,DVO3,DVO4,DVO5,DVO6} I slightly went from
this BWW-type theory and advocated co-existence of two Weinberg's
equations with opposite signs in the mass term for the spin $j=1$ case.
Their connections with classical and quantum electrodynamics, and (the
paper in preparation) the possibility of the use of the same $(j,0)\oplus
(0,j)$ field operator to obtain the Majorana states or the Dirac states
were studied.
The reason for this reformulation is that the Weinberg equations are of
the second order in derivatives and each of them provides dispersional
relationes with both positive and negative signs of the
energy.\footnote{Generally speaking, the both massive Weinberg's equations
possess tachyonic solutions. While this content is {\it not} already in a
strong contradiction with experimental observations (see the enumeration
of experiments on superluminal phenomena above and the papers of E.
Recami) someone can still regard this as a shortcoming before that time
when they would be given adequate explanation.  We note that one can still
escape from this problem by choosing the particular $a$ and $b$ in the
equation below:  \begin{equation} \left [\gamma_{\alpha\beta} p_\alpha
p_\beta + a p_\alpha p_\alpha + b m^2\right ] \psi =0\quad.\label{gen}
\end{equation} Thus, one can obtain the Hammer-Tucker
equations~\cite{HAMMER}, which have dispersion relations $E=\pm \sqrt{{\bf
p}^{\,2} +m^2}$ if one restricts by particles with mass.} Below I
present a brief content of those papers.

The $2(2j+1)$- component analogues of the Dirac functions in the momentum
space were earlier defined as
\begin{equation}\label{pos} {\cal U} ({\bf
p})= {m\over \sqrt{2}} \left (\matrix{ D^J \left (\alpha({\bf p})\right
)\xi_\sigma\cr D^J \left (\alpha^{-1\,\dagger}({\bf p})\right
)\xi_\sigma\cr }\right )\quad, \end{equation} for positive-energy states,
and \begin{equation}\label{neg} {\cal V} ({\bf p})= {m\over \sqrt{2}}
\left (\matrix{ D^J \left (\alpha({\bf p})\Theta_{[1/2]}\right
)\xi^*_\sigma\cr D^J \left ( \alpha^{-1\,\dagger}({\bf p})
\Theta_{[1/2]}\right )(-1)^{2J}\xi^*_\sigma\cr }\right )\quad,
\end{equation} for
negative-energy states, {\it e.g.}, ref.~\cite{NOVOZH}. The following
notation was used
\begin{equation}
\alpha({\bf p})=\frac{p_0+m+(\bbox{\sigma}
\cdot{\bf p})}{\sqrt{2m(p_0+m)}},\quad
\Theta_{[1/2]}=-i\sigma_2\quad.
\end{equation}
For example, in the case of spin $j=1$, one has
\begin{mathletters}
\begin{eqnarray}
&&D^{\,1}\left (\alpha({\bf p})\right ) \,=\,
1+\frac{({\bf J}\cdot{\bf p})}{m}+
\frac{({\bf J}\cdot{\bf p})^2}{m(p_0+m)}\quad,\\
&&D^{\,1}\left (\alpha^{-1\,\dagger}({\bf p})\right ) \,=\,
1-\frac{({\bf J}\cdot{\bf p})}{m}+
\frac{({\bf J}\cdot{\bf p})^2}{m(p_0+m)}\quad,  \\
&&D^{\,1}\left (\alpha({\bf p}) \Theta_{[1/2]}\right ) \,=\,
\left [1+\frac{({\bf J}\cdot{\bf p})}{m}+
\frac{({\bf J}\cdot{\bf p})^2}{m(p_0+m)}\right ]\Theta_{[1]}\quad, \\
&&D^{\,1}\left (\alpha^{-1\,\dagger}({\bf p}) \Theta_{[1/2]}\right ) \,=\,
\left [1-\frac{({\bf J}\cdot{\bf p})}{m}+
\frac{({\bf J}\cdot{\bf p})^2}{m(p_0+m)}\right ]\Theta_{[1]}\quad;
\end{eqnarray}
\end{mathletters}
$\Theta_{[1/2]}$,\,$\Theta_{[1]}$ are the Wigner time-reversal operators
for spin 1/2 and 1, respectively. These definitions lead to the
formulation in which the physical content given by
positive and negative-energy `bispinors'
is the same (like in the paper
of R. H. Tucker and C. L. Hammer~\cite{HAMMER}). One can
consider that
${\cal V}_\sigma ({\bf p}) = (-1)^{1-\sigma}\gamma_5
S^c_{[1]} {\cal U}_{-\sigma} ({\bf p})$ and, thus, the explicit
form of the negative-energy solutions would be the same as of the
positive-energy solutions in accordance with definitions
(\ref{pos},\ref{neg}).

Next, let me look at the Proca equations for a $j=1$ massive particle
\begin{mathletters}
\begin{eqnarray}\label{eq:01}
&&\partial_\mu F_{\mu\nu} = m^2 A_\nu \quad, \\
\label{eq:02}
&&F_{\mu\nu} = \partial_\mu A_\nu - \partial_\nu A_\mu
\end{eqnarray}
\end{mathletters}
in the form given by refs.~\cite{SANKAR,LURIE}.
The Euclidean metric,
$x_\mu =  (\vec x, x_4 =it)$ and notation
$\partial_\mu = (\vec \nabla, -i\partial/\partial t)$,
$\partial_\mu^{\,2} = \vec \nabla^{\,2} -\partial_t^2$, are
used. By means of the choice of  $F_{\mu\nu}$ components
as physical variables one can rewrite the set of equations to
\begin{equation}\label{eq:eq}
m^2 F_{\mu\nu} =\partial_\mu \partial_\alpha F_{\alpha\nu}
-\partial_\nu \partial_\alpha F_{\alpha\mu}
\end{equation}
and
\begin{equation}\label{eq:2}
\partial_\lambda^2 F_{\mu\nu} = m^2 F_{\mu\nu}\quad.
\end{equation}
It is easy to show that they can be represented in the form
($F_{44}=0$, $F_{4i} =i E_i$ and $F_{jk} =\epsilon_{jki} B_i$;\,
$p_\alpha=-i\partial_\alpha$):
\begin{eqnarray}\label{eq:aux1}
\cases{(m^2 +p_4^2) E_i +p_i p_j E_j +
i\epsilon_{ijk} p_4 p_j B_k=0& \cr
&\cr
(m^2 +\vec p^{\,2}) B_i -p_i p_j B_j +
i\epsilon_{ijk} p_4 p_j E_k =0\quad, &}
\end{eqnarray}
or
\begin{eqnarray}\label{eq:aux}
\cases{\left [ m^2 +p_4^2 +{\bf p}^{\,2} -
({\bf J}\cdot {\bf p})^2 \right ]_{ij}
E_j +p_4 ({\bf J}\cdot {\bf p})_{ij} B_j = 0&\cr
&\cr
\left [ m^2 +({\bf J}\cdot {\bf p})^2 \right ]_{ij} B_j +
p_4 ({\bf J}\cdot {\bf p})_{ij} E_j =0\quad. &}
\end{eqnarray}
After adding and subtracting the  obtained equations yield
\begin{eqnarray}
\cases{m^2 ({\bf E} +i{\bf B})_i + p_\alpha p_\alpha {\bf E}_i
- ({\bf J}\cdot {\bf p})^2_{ij} ({\bf E} -i {\bf B})_j
+ p_4 ({\bf J}\cdot {\bf p})_{ij} ({\bf B} +i{\bf E})_j = 0 &\cr
&\cr
m^2 ({\bf E} - i{\bf B})_i + p_\alpha p_\alpha {\bf E}_i
- ({\bf J}\cdot{\bf p})^2_{ij}
({\bf E} +i{\bf B})_j + p_4 ({\bf J}\cdot {\bf p})_{ij}
({\bf B} -i{\bf E})_j = 0\quad, &}
\end{eqnarray}
with $({\bf J}_i)_{jk} = -i\epsilon_{ijk}$ being
the $j=1$ spin matrices.
Equations are equivalent (within a constant factor) to
the Hammer-Tucker equation~\cite{HAMMER}
\begin{equation}\label{eq:Tucker}
(\gamma_{\alpha\beta}p_\alpha p_\beta
+p_\alpha p_\alpha +2 m^2 ) \psi_1 =0 \quad ,
\end{equation}
in the case of the choice $\chi= {\bf E} +i{\bf B}$
and $\varphi ={\bf E} -i{\bf B}$,\,\,\,
$\psi_1 = \mbox{column}\, (\chi , \quad \varphi)$. Matrices
$\gamma_{\alpha\beta}$ are the
covariantly defined matrices of Barut,
Muzinich and Williams~\cite{BARUT0} for spin $j=1$.
The equation (\ref{eq:Tucker}) for massive particles is
characterized by positive- and negative-energy solutions with a physical
dispersion only $E_p =\pm \sqrt{{\bf p}^{\,2} + m^2}$, the determinant is
equal to
\begin{equation}
\mbox{Det} \left [\gamma_{\alpha\beta} p_\alpha p_\beta
+p_\alpha p_\alpha +2m^2 \right ] = -64 m^6 (p_0^2 -{\bf p}^{\,2}
-m^2)^3\quad, \label{det}
\end{equation}
However, there is another
equation which also do not have `acausal' solutions.  The second one (with
$a=-1$ and $b=-2$, see (\ref{gen})) is \begin{equation}
(\gamma_{\alpha\beta}p_\alpha p_\beta - p_\alpha p_\alpha - 2m^2) \psi_2
=0 \quad .  \label{eq:Tucker2}
\end{equation}
In the tensor form it leads to the equations
which are dual  to (\ref{eq:aux1})
\begin{eqnarray}
\cases{(m^2 +{\bf p}^{\,2}) C_i - p_i
p_j C_j - i\epsilon_{ijk} p_4 p_j D_k = 0 & \cr
&\cr
(m^2 +p_4^2 )D_i + p_i p_j
D_j - i\epsilon_{ijk} p_4 p_j C_k =0\quad. &}
\end{eqnarray}
They can be
rewritten in the form, {\it cf.} (\ref{eq:eq}),
\begin{equation}\label{eq:eqd}
m^2 \widetilde F_{\mu\nu} =\partial_\mu \partial_\alpha \widetilde
F_{\alpha\nu}
-\partial_\nu \partial_\alpha \widetilde F_{\alpha\mu} \quad ,
\end{equation}
with $\widetilde F_{4i} = iD_i$ and $\widetilde F_{jk} =
- \epsilon_{jki} C_i$.
The vector $C_i$ is an analog of $E_i$ and $D_i$ is an analog of $B_i$
because in some cases it is convenient to equate
$\widetilde F_{\mu\nu} ={1\over 2} \epsilon_{\mu\nu\rho\sigma}
F_{\rho\sigma}$, $\epsilon_{1234} = -i$.
The following
properties of the antisymmetric Levi-Civita tensor
$$ \epsilon_{ijk}
\epsilon_{ijl} =2\delta_{kl}\quad, \quad \epsilon_{ijk} \epsilon_{ilm} =
(\delta_{jl} \delta_{km} -\delta_{jm} \delta_{kl} )\quad,$$
and
$$ \epsilon_{ijk}
\epsilon_{lmn} =  \mbox{Det}\, \pmatrix{\delta_{il} & \delta_{im} &
\delta_{in}\cr
\delta_{jl} & \delta_{jm} & \delta_{jn}\cr
\delta_{kl} & \delta_{km} & \delta_{kn}} \quad $$
have been used.

Comparing the structure of the Weinberg equation ($a=0$, $b=1$)
with the Hammer-Tucker `doubles' one can convince ourselves
that the former can be represented in the tensor form:
\begin{equation}\label{eq:3}
m^2 F_{\mu\nu} =\partial_\mu \partial_\alpha F_{\alpha\nu}
-\partial_\nu \partial_\alpha F_{\alpha\mu} + {1\over 2}
(m^2 - \partial_\lambda^2) F_{\mu\nu}\quad,
\end{equation}
that corresponds to Eq. (\ref{w1}).
However, as we learned, it is possible to build an equation --- `double'  :
\begin{equation}\label{eq:4}
m^2 \widetilde F_{\mu\nu} =\partial_\mu \partial_\alpha \widetilde
F_{\alpha\nu}
-\partial_\nu \partial_\alpha \widetilde F_{\alpha\mu} +
{1\over 2} (m^2 -\partial_\lambda^2) \widetilde F_{\mu\nu}\quad,
\end{equation}
that corresponds to Eq. (\ref{w2}).
The Weinberg's set of equations is written in the form:
\begin{mathletters}
\begin{eqnarray}\label{w1}
(\gamma_{\alpha\beta} p_\alpha p_\beta + m^2 )\psi_1 &=& 0\quad,\\
\label{w2}
(\gamma_{\alpha\beta} p_\alpha p_\beta - m^2) \psi_2 &=& 0 \quad.
\end{eqnarray}
\end{mathletters}
Thanks to the Klein-Gordon equation (\ref{eq:2}) these equations
are equivalent to the Proca tensor equations (\ref{eq:eq},\ref{eq:eqd}),
and to the Hammer-Tucker `doubles', in a free case.  However, if an
interaction is included, one cannot say that.  The second equation
(\ref{w2}) coincides with the Ahluwalia {\it et al.} equation  for $v$
spinors (Eq.  (16) of ref.~[3b]) or with Eq.  (12) of ref.~[62b].  Thus,
the general solution describing $j=1$ states can be presented as a
superposition \begin{equation}\label{eq:super} \Psi^{(1)} = c_1
\psi_1^{(1)} + c_2 \psi_2^{(1)} \quad, \end{equation} where the constants
$c_1$ and $c_2$ are to be defined from the boundary, initial and
normalization conditions.  Let me note a surprising fact:  while both the
massive Proca equations (or the Hammer-Tucker ones) and the Klein-Gordon
equation do not possess `non-physical' solutions, their sum, Eqs.
(\ref{eq:3},\ref{eq:4}), or the Weinberg equations (\ref{w1},\ref{w2}),
acquire tachyonic solutions.  Next, equations (\ref{w1}) and (\ref{w2})
can recast in another form (index $``T"$ denotes a transpose matrix):
\begin{mathletters}
\begin{eqnarray}\label{w11}
\left [\gamma_{44} p_4^2 +2\gamma_{4i}^{^T} p_4 p_i +\gamma_{ij}p_i p_j
-m^2\right ] \psi_1^{(2)} &=&0 \quad ,\\ \label{w21}
\left [\gamma_{44} p_4^2 +2\gamma_{4i}^{^T} p_4 p_i +\gamma_{ij}p_i p_j
+m^2 \right ] \psi_2^{(2)} &=&0 \quad ,
\end{eqnarray}
\end{mathletters}
respectively, if understand
$\psi_1^{(2)} \sim \mbox{column} \, (B_i + iE_i ,\quad B_i -iE_i)
=i\gamma_5 \gamma_{44} \psi_1^{(1)}$ and
$\psi_2^{(2)} \sim \mbox{column} \, (D_i + i
C_i, \quad D_i - iC_i )= i\gamma_5 \gamma_{44} \psi_2^{(1)}$.
The general solution is again a linear combination
\begin{equation}
\Psi^{(2)} =c_1 \psi_1^{(2)} + c_2 \psi_2^{(2)}\quad.
\end{equation}

From, {\it e.g.}, Eq. (\ref{w1}),
dividing $\psi^{(1)}_1$
into longitudinal and transversal parts one can come to
the equations
\begin{eqnarray}
\lefteqn{\left [E^2 -\vec p^{\,2}\right ]({\bf
E}+i{\bf B})^{\parallel} -m^2 ({\bf E}-i{\bf B})^{\parallel} +\nonumber}\\
&+&\left [E^2 +\vec p^{\,2}- 2 E ({\bf J}\cdot{\bf p})\right
] ({\bf E}+i{\bf B})^{\perp} - m^2 ({\bf E}-i{\bf B})^{\perp}
=0\quad,\label{cl1}
\end{eqnarray} and
\begin{eqnarray} \lefteqn{\left
[E^2 -\vec p^{\,2}\right ]({\bf E}-i{\bf B})^{\parallel} -m^2
({\bf E}+i{\bf B})^{\parallel} +\nonumber}\\ &+&\left [E^2 + \vec
p^{\,2}+2E ({\bf J}\cdot{\bf p})\right ]({\bf E}-i{\bf B})^{\perp}
- m^2 ({\bf E}+i{\bf B})^{\perp} =0 \quad.\label{cl2} \end{eqnarray}
One can see that in the classical field theory antisymmetric tensor matter
fields are the fields having transversal components in the massless limit.
Under the transformations $\psi_1^{(1)} \rightarrow \gamma_5 \psi_2^{(1)}$
or $\psi_1^{(2)} \rightarrow \gamma_5 \psi_2^{(2)}$ the set of equations
(\ref{w1}) and (\ref{w2}), or Eqs. (\ref{w11}) and (\ref{w21}),  leave to
be invariant.  The origin of this fact is the dual invariance of the set
of the Proca equations.  In the matrix form dual transformations
correspond to the chiral transformations.

Let me consider the question of the `double' solutions
on the basis of spinorial analysis. In ref.~[62a,p.1305]
(see also~\cite[p.60-61]{LANDAU2})
relations between the Weinberg $j=1$ ``bispinor" (bivector, indeed)
and symmetric spinors of $2j$- rank have been discussed.
It was noted there: ``The wave
function may be written in terms of two
three-component functions $\psi=\mbox{column}
(\chi \quad \varphi)$,
that, for the continuous group, transform independently
each of other and that are related to two symmetric
spinors:
\begin{mathletters}
\begin{eqnarray}
&&\chi_1 = \chi_{\dot 1\dot 1} , \quad \chi_2 = \sqrt{2}
\chi_{\dot 1 \dot 2} , \quad \chi_3 =
\chi_{\dot 2 \dot 2} \quad,\\
&& \varphi_1 = \varphi^{11} , \quad \varphi_2 = \sqrt{2}
\varphi^{12} , \quad \varphi_3 = \varphi^{22} \quad,
\end{eqnarray}
\end{mathletters}
when the standard representation for the spin-one
matrices, with $J_3$ diagonal is used."
Under the inversion operation we
have the following rules~\cite[p.59]{LANDAU2}:
$\varphi^\alpha \rightarrow \chi_{\dot\alpha}$,\,
$\chi_{\dot
\alpha} \rightarrow \varphi^{\alpha}$,\,
$\varphi_\alpha \rightarrow -\chi^{\dot \alpha}$
and $\chi^{\dot\alpha}\rightarrow -\varphi_\alpha$.
Hence, one can deduce (if one understand $\chi_{\dot\alpha\dot\beta}
=\chi_{\{\dot\alpha} \chi_{\dot\beta\}}$\, ,\, $\varphi^{\alpha\beta}
=\varphi^{\{\alpha}\varphi^{\beta\}}$)
\begin{mathletters}
\begin{eqnarray}
&&\chi_{\dot 1 \dot 1} \rightarrow \varphi^{11} \quad, \quad
\chi_{\dot 2 \dot 2} \rightarrow \varphi^{22} \quad, \quad
\chi_{ \{ \dot 1 \dot 2 \} } \rightarrow
\varphi^{ \{ 12 \} } \quad,\\
&& \varphi^{11} \rightarrow \chi_{\dot 1 \dot 1} \quad, \quad
\varphi^{22} \rightarrow \chi_{\dot 2 \dot 2} \quad, \quad
\varphi^{ \{ 12 \} } \rightarrow \chi_{ \{ \dot 1 \dot 2 \} } \quad.
\end{eqnarray}
\end{mathletters}
However, this definition of symmetric spinors
of the second rank $\chi$ and $\varphi$ is ambiguous.
We are also able to define, {\it e.g.}, $\tilde \chi_{\dot\alpha\dot\beta}
=\chi_{\{\dot\alpha} H_{\dot\beta\}}$ and
$\tilde \varphi^{\alpha\beta} = \varphi^{\{\alpha}
\Phi^{\beta\}}$,
where $H_{\dot\beta} = \varphi_{\beta}^{*}$,
$\Phi^{\beta} =(\chi^{\dot\beta})^{*}$.
It is straightforwardly showed that in the framework
of the second definition we
have under the space-inversion operation:
\begin{mathletters}
\begin{eqnarray}
&&\tilde \chi_{\dot 1 \dot 1} \rightarrow
-\tilde \varphi^{11} \quad , \quad
\tilde \chi_{\dot 2 \dot 2} \rightarrow
- \tilde \varphi^{22} \quad , \quad
\tilde \chi_{ \{ \dot 1 \dot 2 \} } \rightarrow
- \tilde \varphi^{ \{ 12 \} } \quad, \\
&&\varphi^{11} \rightarrow -\tilde \chi_{\dot 1 \dot 1}\quad , \quad
\tilde \varphi^{22} \rightarrow -\tilde \chi_{\dot 2
\dot 2}\quad , \quad
\tilde \varphi^{ \{ 12 \} } \rightarrow
-\tilde \chi_{ \{ \dot 1 \dot 2 \} } \quad .
\end{eqnarray}
\end{mathletters}
The Weinberg `bispinor'
$(\chi_{\dot\alpha\dot\beta} \quad \varphi^{\alpha\beta})$
corresponds to the equations (\ref{w11}) and (\ref{w21}) ,
meanwhile
$(\tilde \chi_{\dot\alpha\dot\beta}\quad
\tilde\varphi^{\alpha\beta})$, to
the equation (\ref{w1}) and (\ref{w2}).
Similar conclusions can be arrived at in
the case of the parity definition
as $P^2 = -1$. Transformation rules are then
$\varphi^\alpha \rightarrow i\chi_{\dot\alpha}$, $\chi_{\dot\alpha}
\rightarrow i\varphi^\alpha$,
$\varphi_\alpha \rightarrow -i\chi^{\dot\alpha}$
and $\chi^{\dot\alpha}\rightarrow -i\varphi_\alpha$,
ref.~\cite[p.59]{LANDAU2} .
Hence, $\chi_{\dot\alpha\dot\beta} \leftrightarrow
-\varphi^{\alpha\beta}$
and $\tilde \chi_{\dot\alpha\dot\beta} \leftrightarrow
- \tilde\varphi^{\alpha\beta}$, but
$\varphi^{\alpha}_{\quad\beta}
\leftrightarrow \chi_{\dot\alpha}^{\quad\dot\beta}$ and $\tilde
\varphi^{\alpha}_{\quad\beta} \leftrightarrow
\tilde \chi_{\dot\alpha}^{\quad\dot\beta}$\quad.

In order to consider the corresponding dynamical content we should choose
an appropriate Lagrangian. In the framework of this essay we concern with
the Lagrangian which is similar to the one used in earlier works on the
$2(2j+1)$ formalism (see for references~\cite{DVOOLD}). Our Lagrangian
includes additional terms which respond to the Weinberg `double' and does
not suffer from the problems noted in the old works. Here it
is:\,\footnote{Under field functions we assume $\psi^{(1)}_{1,2}$. Of
course, one can use another form with substitutions:  $\psi_{1,2}^{(1)}
\rightarrow \psi_{2,1}^{(2)}$ and $\gamma_{\mu\nu} \rightarrow \widetilde
\gamma_{\mu\nu}$, where $\widetilde \gamma_{\mu\nu} \equiv
\gamma_{\mu\nu}^{^T} \equiv \gamma_{44}
\gamma_{\mu\nu} \gamma_{44}$.}\, $^{,}$\,
\footnote{Questions related
with other possible Lagrangians will be solved in other papers.  The first
alterantive form is with the following dynamical part:
\begin{equation} {\cal
L}^{(2^\prime)} = - \partial_\mu \overline \psi_1 \gamma_{\mu\nu}
\partial_\nu \psi_2 -\partial_\mu \overline \psi_2 \gamma_{\mu\nu}
\partial_\nu \psi_1  \quad, \nonumber \end{equation} where $\psi_1$ and
$\psi_2$ are defined by the equations (\ref{w1},\ref{w2}).
This form appear not to admit the mass term in an ordinary sense.
The second Lagrangian composed of the two states
($\psi_1$, the solution of the equation (\ref{w1})
and $\psi_2$, of the equation (\ref{w21}), {\it e.g.},
\begin{equation}\label{eq:Lagran2} {\cal
L}^{(2^{\prime\prime})} = - \partial_\mu \overline \psi_1 \widetilde
\gamma_{\mu\nu} \partial_\nu \psi_2 - \partial_\mu \overline \psi_2
\gamma_{\mu\nu} \partial_\nu \psi_1 - m^2 \overline \psi_2 \psi_1 - m^2
\overline \psi_1 \psi_2 \quad
\end{equation}
is not Hermitian.}
\begin{equation}\label{eq:Lagran1}
{\cal L} = -\partial_\mu \overline
\psi_1  \gamma_{\mu\nu} \partial_\nu \psi_1 -\partial_\mu \overline \psi_2
\gamma_{\mu\nu} \partial_\nu \psi_2 - m^2 \overline \psi_1 \psi_1 + m^2
\overline \psi_2 \psi_2 \quad .
\end{equation}
The Lagrangian
(\ref{eq:Lagran1}) leads to the equations (\ref{w1},\ref{w2}) which
possess solutions with  a ``correct" (bradyon) physical dispersion
and tachyonic solutions as well.
This Lagrangian (\ref{eq:Lagran1}) is scalar, Hermitian and it contains
only first-order time derivatives.  In order to obtain Lagrangians
corresponding to the Tucker-Hammer set (\ref{eq:Tucker},\ref{eq:Tucker2}),
obviously, one should  add in (\ref{eq:Lagran1}) terms answering for the
Klein-Gordon equation.

At this point  I would like to regard the question of solutions
in the momentum space. Using the plane-wave expansion
\begin{mathletters}
\begin{eqnarray}\label{pl1}
\psi_1 (x) &=&\sum_\sigma \int \frac{d^3 {\bf p}}{(2\pi)^3} \frac{1}{m
\sqrt{2E_p}} \left [ {\cal U}_1^{\,\sigma} ({\bf p}) a_\sigma
({\bf p}) e^{ip\cdot x} +{\cal V}_1^{\,\sigma} ({\bf p})
b^{\,\dagger}_\sigma ({\bf p}) e^{-ip\cdot x} \right ]\quad,\\ \label{pl2}
\psi_2 (x) &=&\sum_\sigma \int \frac{d^3 {\bf p}}{(2\pi)^3}
\frac{1}{m\sqrt{2E_p}} \left [ {\cal U}_2^{\,\sigma} ({\bf p})
c_\sigma ({\bf p}) e^{ip\cdot x} +{\cal V}_2^{\,\sigma} ({\bf p})
d^{\,\dagger}_\sigma ({\bf p}) e^{-ip\cdot x} \right ] \quad,
\end{eqnarray}
\end{mathletters}
one can see that the momentum-space
`double' equations
\begin{mathletters}
\begin{eqnarray}\label{eq:sp1} \left [ - \gamma_{44}
E^2 +2iE\gamma_{4i} {\bf p}_i +\gamma_{ij} {\bf p}_i {\bf p}_j  +
m^2\right ] {\cal U}_1^\sigma ({\bf p}) &=& 0 \quad (\mbox{or}\,\,{\cal
V}_1^\sigma ({\bf p})) \quad, \\ \label{eq:sp2} \left [ - \gamma_{44} E^2
+2iE\gamma_{4i} {\bf p}_i +\gamma_{ij} {\bf p}_i {\bf p}_j  - m^2\right ]
{\cal U}_2^\sigma ({\bf p}) &=& 0 \quad (\mbox{or} \,\,{\cal V}_2^\sigma
({\bf p}))\quad \end{eqnarray}
\end{mathletters}
are satisfied by bispinors
\begin{eqnarray}\label{bb1}
{\cal U}_1^{(1)\,\sigma} ({\bf p})=
\frac{m}{\sqrt{2}}\pmatrix{\left [ 1+ {({\bf J}\cdot{\bf p})\over  m}
+{({\bf J}\cdot {\bf p})^2 \over  m(E_p +m)}\right ]\xi_\sigma \cr \left [
1 - {({\bf J}\cdot{\bf p})\over  m} +{({\bf J}\cdot {\bf p})^2 \over
m(E_p + m)}\right ] \xi_\sigma\cr}\quad, \end{eqnarray}
and
\begin{eqnarray}\label{bb2} {\cal U}_2^{(1)\,\sigma} ({\bf p})
=\frac{m}{\sqrt{2}}\pmatrix{\left [ 1+ {({\bf J}\cdot{\bf p})\over  m} +
{({\bf J}\cdot {\bf p})^2 \over  m(E_p +m)}\right ] \xi_{\sigma} \cr \left
[ - 1 + {({\bf J}\cdot{\bf p})\over  m} - {({\bf J}\cdot {\bf p})^2 \over
m(E_p+ m)}\right ] \xi_{\sigma}\cr}\quad,
\end{eqnarray}
respectively.
The form (\ref{bb1}) has been presented by Hammer, Tucker and Novozhilov
in refs.~\cite{HAMMER,NOVOZH}.  The bispinor
normalization in most of the previous papers is chosen to unit. However,
as mentioned in ref.~\cite{DVA1} it is more convenient to work with
bispinors normalized to the mass, {\it e.g.}, $\pm m^{2j}$ in order to
make zero-momentum spinors to vanish in the massless limit.  Here and
below I keep the normalization of bispinors as in ref.~\cite{DVA1}.
Bispinors of Ahluwalia {\it et al.}, ref.~\cite{DVA1}, can be written in
the  more compact  form:  \begin{eqnarray} u^{\sigma}_{AJG} ({\bf p}) =
\pmatrix{\left [ m +\frac{({\bf J}\cdot {\bf p})^2}{E_p+m} \right ]
\xi_{\sigma}\cr ({\bf J}\cdot {\bf p}) \xi_{\sigma}\cr}\quad, \quad
v^{\sigma}_{AJG} ({\bf p}) =\pmatrix{0 & 1\cr 1 & 0\cr} u^{\sigma}_{AJG}
({\bf p})\quad.  \end{eqnarray}
They coincide with the Hammer-Tucker-Novozhilov bispinors within a
normalization and a unitary transformation by ${\bf U}$ matrix:
\begin{mathletters}
\begin{eqnarray}
u^{\sigma}_{~\cite{DVA1}} ({\bf p}) &=& m\,\,\, \cdot  {\bf U}\,
{\cal U}^{\sigma}_{~\cite{HAMMER,NOVOZH}} ({\bf p}) =
\frac{m}{\sqrt{2}}\pmatrix{1 & 1 \cr 1 & -1\cr}
{\cal U}^{\sigma}_{~\cite{HAMMER,NOVOZH}} ({\bf p})\quad,\\
v^{\sigma}_{~\cite{DVA1}} ({\bf p}) &=& m\,\,\, \cdot {\bf U} \,
\gamma_{5} {\cal U}^{\sigma}_{~\cite{HAMMER,NOVOZH}} ({\bf p}) =
\frac{m}{\sqrt{2}}\pmatrix{1 & 1 \cr 1 & -1\cr} \gamma_{5}
{\cal U}^{\sigma}_{~\cite{HAMMER,NOVOZH}} ({\bf p}) \quad.
\end{eqnarray}
\end{mathletters}
But, as we found the Weinberg equations
(with $+m^2$ and with $-m^2$)  have solutions with both positive- and
negative-energies. We have to propose the interpretation of the latter.
In the framework of this paper one can consider that
\begin{equation}
{\cal V}^{(1)}_\sigma ({\bf p}) = (-1)^{1-\sigma}\gamma_5 S^c_{[1]}
{\cal U}^{(1)}_{-\sigma} ({\bf p}) \label{negex}
\end{equation}
and, thus, the explicit forms of the
negative-energy solutions would be the same as those of the positive-energy
solutions in accordance with definitions (\ref{pos},\ref{neg}).
Thus, in
the case of the choice ${\cal U}_1^{(1)\,\sigma} ({\bf p})$ and
${\cal V}_2^{(1)\,\sigma} ({\bf p}) \sim \gamma_5 {\cal U}_1^{(1)\,\sigma}
({\bf p})$ as physical bispinors we come to the
Bargmann-Wightman-Wigner-type (BWW) quantum field model proposed by
Ah\-lu\-wa\-lia {\it et al.} Of course, following  the same logic one
can choose ${\cal U}_2^{(1)\,\sigma}$ and ${\cal V}_1^{(1)\,\sigma}$
as positive- and negative-
bispinors, respectively, and come to a reformulation of the BWW
theory.  While in this case parities of a boson and its antiboson are
opposite, we have $-1$ for ${\cal U}$- bispinor and $+1$ for ${\cal V}$-
bispinor, {\it i.e.} different in the sign from the model of Ahluwalia
{\it et al.}\footnote{At the present level of our knowledge this
mathematical difference has no physical significance, but we want to stay
at the most general positions. Perhaps, some yet unknown forms of
interactions ({\it e.g.} of neutral particles) can lead to the observed
physical difference between these models.} In the meantime, the construct
proposed by Weinberg~\cite{WEIN1} and developed in this paper is also
possible. The ${\cal V}^{(1)\,\sigma}_1 (\vec p)$ as defined by
(\ref{negex}) can also be solutions of the equation (\ref{w1}).  The
origin of the possibility that the positive- and negative-energy solutions
in Eqs.  (\ref{eq:sp1},\ref{eq:sp2}) can coincide each other is the
following:  the Weinberg equations are of the second order in  time
derivatives.  The Bargmann-Wightman-Wigner construct presented by
Ahluwalia~\cite{DVA1} is not the only construct in the $(1,0)\oplus (0,1)$
representation and one can start with the earlier definitions of the
$2(2j+1)$ bispinors.

Next, previously we gave two additional equations
(\ref{w11},\ref{w21}). Their solutions can  also be useful because
of the possibility of the use of different Lagrangian forms.
Solutions in the momentum representation are
written
\begin{eqnarray}\label{b11} {\cal U}_1^{(2)\,\sigma}
({\bf p})= \frac{m}{\sqrt{2}}\pmatrix{\left [1- {({\bf J}\cdot{\bf
p})\over m} +{({\bf J}\cdot {\bf p})^2 \over m (E_p+m)}\right ]\xi_\sigma
\cr \left [ -1  - {({\bf J}\cdot{\bf p})\over m}  -{({\bf J}\cdot {\bf
p})^2 \over m (E_p + m)}\right ] \xi_\sigma}\quad,
\end{eqnarray}
\begin{eqnarray}\label{b21} {\cal U}_2^{(2)\,\sigma} ({\bf p})=
=\frac{m}{\sqrt{2}}\pmatrix{\left [1-
{({\bf J}\cdot{\bf p}) \over m} + {({\bf J}\cdot {\bf p})^2 \over m (E_p +
m)}\right ]\xi_\sigma \cr \left [ 1  + {({\bf J}\cdot{\bf p}) \over m} +
{({\bf J}\cdot {\bf p})^2 \over  m (E_p + m)}\right ] \xi_\sigma}\quad.
\end{eqnarray} Therefore,  one has ${\cal U}_2^{(1)} ({\bf p}) = \gamma_5
{\cal U}_1^{(1)} ({\bf p})$ and $\overline {\cal U}_2^{(1)} ({\bf p}) =
-\overline {\cal U}_1^{(1)} ({\bf p})\gamma_5$; \, ${\cal U}_1^{(2)}
({\bf p}) = \gamma_5\gamma_{44} {\cal U}_1^{(1)} ({\bf p})$ and $\overline
{\cal U}_1^{(2)} = \overline {\cal U}_1^{(1)} \gamma_5\gamma_{44}$; ${\cal
U}_2^{(2)} ({\bf p}) = \gamma_{44} {\cal U}_1^{(1)} ({\bf p})$ and
$\overline {\cal U}_2^{(2)} ({\bf p}) =  \overline {\cal
U}_1^{(1)}\gamma_{44}$.  In fact, they are connected by the
transformations of the inversion group.

Let me now apply the quantization procedure to the Weinberg fields.
From the definitions~\cite{LURIE}:
\begin{mathletters}
\begin{eqnarray}
{\cal T}_{\mu\nu} &=& -\sum_i \left \{
\frac{\partial {\cal L}}{\partial (\partial_\mu \phi_i )}
\partial_\nu \phi_i
+ \partial_\nu \overline \phi_i\frac{\partial {\cal L}}{\partial
(\partial_\mu \overline \phi_i )} \right \}
+{\cal L}\delta_{\mu\nu}\quad,\\
P_\mu &=& \int {\cal P}_{\mu} (x) d^3 x = -i\int {\cal T}_{4\mu} d^3 x
\end{eqnarray}
\end{mathletters}
one can find  the energy-momentum tensor
\begin{eqnarray}
\lefteqn{{\cal T}_{\mu\nu} = \partial_\alpha \overline \psi_1
\gamma_{\alpha\mu} \partial_\nu \psi_1 + \partial_\nu \overline \psi_1
\gamma_{\mu\alpha} \partial_\alpha \psi_1 +\nonumber}\\ &+&\partial_\alpha
\overline \psi_2 \gamma_{\alpha\mu} \partial_\nu \psi_2 +\partial_\nu
\overline \psi_2 \gamma_{\mu\alpha} \partial_\alpha \psi_2+{\cal
L}\delta_{\mu\nu} \quad.
\end{eqnarray}
As a result the Hamiltonian is
\begin{eqnarray}\label{H}
\lefteqn{{\cal H} = \int \left [ -\partial_4 \overline \psi_2
\gamma_{4 4}\partial_4 \psi_2 + \partial_i \overline \psi_2  \gamma_{ij}
\partial_j \psi_2 -\right.\nonumber}\\ &-& \left.  \partial_4 \overline
\psi_1 \gamma_{4 4} \partial_4 \psi_1 + \partial_i \overline \psi_1
\gamma_{ij} \partial_j \psi_1 +m^2 \overline \psi_1 \psi_1 -m^2 \overline
\psi_2 \psi_2 \right ] d^3 x \quad.
\end{eqnarray}
Using the plane-wave expansion and the procedure of, {\it e.g.},
ref.~\cite{BOGOLU} one can come to the quantized Hamiltonian
\begin{equation}
{\cal H} =  \sum_\sigma \int \frac{d^3 {\bf p}}{(2\pi)^3} E_p \,
\left [ a_\sigma^{\,\dagger} ({\bf p}) a_\sigma ({\bf p})
+b_\sigma ({\bf p}) b_\sigma^{\,\dagger} ({\bf p})
+c_\sigma^{\,\dagger} ({\bf p}) c_\sigma({\bf p}) +d_\sigma
({\bf p}) d^{\,\dagger}_\sigma ({\bf p})\right ]\quad,
\end{equation}
Therefore, following the
standard textbooks,{\it e.g.}, refs.~\cite{BOGOLU,LANDAU2}
the commutation relations can be set up as follows:
\begin{mathletters}
\begin{eqnarray}\label{c1} \left [a_\sigma ({\bf p}),
a^\dagger_{\sigma^\prime} ({\bf k}) \right ]_{-} &=& \left [c_\sigma ({\bf
p}), c^\dagger_{\sigma^\prime} ({\bf k}) \right ]_{-} = (2\pi)^3
\delta_{\sigma\sigma^\prime}\delta ({\bf p} - {\bf k})\quad,\\ \label{c2}
\left [b_\sigma ({\bf p}), b^\dagger_{\sigma^\prime}
({\bf k}) \right ]_{-} &=&  \left [d_\sigma ({\bf p}),
d^\dagger_{\sigma^\prime} ({\bf k}) \right ]_{-} =
(2\pi)^3
\delta_{\sigma\sigma^\prime}\delta ({\bf p} - {\bf k})\quad,
\end{eqnarray}
\end{mathletters}
or even in the more general form~\cite{DVO2,DVO6}.
It is easy to see that the Hamiltonian is  positive-definite and
the translational invariance still keeps  in the framework of this
description ({\it cf.} with refs.~\cite{WEIN1,DVA1}). Please pay attention
here:  {\it I did never apply the indefinite metric}.

Analogously, from the definitions
\begin{mathletters}
\begin{eqnarray}
{\cal J}_{\mu} &=& -i \sum_i \left \{ \frac{\partial {\cal L}}{\partial
(\partial_\mu \phi_i)}
\phi_i -\overline \phi_i \frac{\partial {\cal L}}{\partial
(\partial_\mu \overline\phi_i)} \right \}\quad, \\
Q &=& -i \int {\cal J}_{4} (x)  d^3 x \quad,
\end{eqnarray}
\end{mathletters}
and
\begin{mathletters}
\begin{eqnarray}\label{PL}
\lefteqn{{\cal M}_{\mu\nu,\lambda} = x_\mu {\cal T}_{\lambda\nu}
- x_\nu {\cal T}_{\lambda\mu} - \nonumber}\\
&-&i \sum_{i}
\left \{ \frac{\partial {\cal L}}
{\partial (\partial_\lambda \phi_i)} N_{\mu\nu}^{\phi_i} \phi_i +
\overline \phi_i  N_{\mu\nu}^{\overline\phi_i} \frac{\partial {\cal L}}
{\partial (\partial_\lambda \overline \phi_i)}\right \} \quad,\\
\label{PL0}
M_{\mu\nu} &=& -i \int  {\cal M}_{\mu\nu, 4} (x) d^3 x  \quad,
\end{eqnarray}
\end{mathletters}
one can find the current operator
\begin{eqnarray}\label{qq1}
\lefteqn{{\cal J}_\mu = i \left [\partial_\alpha \overline \psi_1
\gamma_{\alpha\mu}
\psi_1 - \overline \psi_1 \gamma_{\mu\alpha} \partial_\alpha  \psi_1
+\right.\nonumber}\\
&+&\left.  \partial_\alpha\overline \psi_2 \gamma_{\alpha\mu}  \psi_2
- \overline \psi_2
\gamma_{\mu\alpha} \partial_\alpha \psi_2\right ]\quad,
\end{eqnarray}
and using (\ref{PL},\ref{PL0})  the spin momentum tensor
\begin{eqnarray}\label{ss1}
\lefteqn{S_{\mu\nu,\lambda} = i \left [ \partial_\alpha \overline \psi_1
\gamma_{\alpha\lambda} N_{\mu\nu}^{\psi_1} \psi_1 +  \overline \psi_1
N_{\mu\nu}^{\overline \psi_1}\gamma_{\lambda\alpha}
\partial_{\alpha} \psi_1 + \right.\nonumber}\\
&+& \left. \partial_\alpha \overline \psi_2
\gamma_{\alpha\lambda} N_{\mu\nu}^{\psi_2} \psi_2
+\overline \psi_2 N_{\mu\nu}^{\overline \psi_2}
\gamma_{\lambda\alpha} \partial_\alpha \psi_2 \right ]\quad.
\end{eqnarray}
If  the Lorentz group generators  (a $j=1$ case) are defined from
\begin{mathletters}
\begin{eqnarray}\label{lor}
&&\overline  \Lambda \gamma_{\mu\nu} \Lambda a_{\mu\alpha}
a_{\nu\beta} = \gamma_{\alpha\beta}\quad,\\
&& \overline \Lambda \Lambda =1\quad,\\ \label{lor3}
&&\overline \Lambda  = \gamma_{44} \Lambda^\dagger \gamma_{44}\quad.
\end{eqnarray}
\end{mathletters}
then in order to keep the  Lorentz covariance of the Weinberg
equations and of the Lagrangian (\ref{eq:Lagran1})
one can use the following generators:
\begin{equation}
N_{\mu\nu}^{\psi_1,  \psi_2 (j=1)} =
- N_{\mu\nu}^{\overline \psi_1 , \overline \psi_2 (j=1)} =
{1\over 6} \gamma_{5,\mu\nu}\quad,
\end{equation}
The matrix  $\gamma_{5,\mu\nu}= i \left [\gamma_{\mu\lambda},
\gamma_{\nu\lambda}\right ]_{-}$
is  defined to be Hermitian.

The quantized charge operator and the quantized  spin
operator follow immediately from (\ref{qq1}) and (\ref{ss1}):
\begin{equation}
Q=\sum_\sigma \int \frac{d^3 {\bf p}}{(2\pi)^3} \, \left [
a_\sigma^\dagger ({\bf p}) a_\sigma ({\bf p}) -
b_\sigma ({\bf p}) b_\sigma^\dagger ({\bf p}) +
c_\sigma^\dagger ({\bf p}) c_\sigma ({\bf p}) -
d_\sigma ({\bf p}) d_\sigma^\dagger ({\bf p})\right ]\quad,
\end{equation}
\begin{eqnarray}\label{spin}
\lefteqn{(W\cdot n) /m=   \sum_{\sigma\sigma^\prime}\int \frac{d^3
{\bf p}}{(2\pi)^3}{1\over m^2 E_p} \overline {\cal U}_1^\sigma ({\bf p})
(E_p \gamma_{44} - i\gamma_{4i}p_i )\,\, I \otimes ({\bf J}\cdot {\bf n})\,
{\cal U}^{\sigma^\prime}_1 ({\bf p}) \times \nonumber}\\ & &\qquad \times
\left [a_\sigma^\dagger ({\bf p}) a_{\sigma^\prime} ({\bf p}) +
c_\sigma^\dagger ({\bf p}) c_{\sigma^\prime} ({\bf p}) - b_\sigma ({\bf
p}) b_{\sigma^\prime}^\dagger ({\bf p}) -d_\sigma  ({\bf p})
d_{\sigma^\prime}^\dagger  ({\bf p}) \right ]\quad
\end{eqnarray}
(provided that  a frame is chosen in such a way
that ${\bf n} \,\,\, \vert\vert \,\,\,{\bf p}$ \,\,\, is along the third
axis).  It is easy to verify  eigenvalues of the charge operator are
$\pm 1$,\footnote{In the Majorana construct the eigenvalue is zero.} and
of the Pauli-Lyuban'sky spin operator are
\begin{equation}\label{heli}
\xi^*_\sigma ({\bf J}\cdot {\bf n}) \xi_{\sigma^\prime} = +1,\, 0\, -1
\end{equation} in a massive case and $\pm 1$ in a massless case (see the
discussion on the massless limit of the Weinberg bispinors in ref.~[3a]).

In order to solve the question of finding propagators in this theory we
would like to consider the most general case. In ref.~[3a]  a particular
case of the BWW bispinors has been regarded. In order to decide what
case is physically relevant for describing one or another situation one
should know what symmetries are respected by the Nature.
So, let me check, if the sum of four equations ($x=x_2 -x_1$)
\begin{eqnarray}\label{prop} &&\hspace*{-1cm}\left [ \gamma_{\mu\nu}
\partial_\mu \partial_\nu -m^2 \right ] \int  \frac{d^3 {\bf p}}{(2\pi)^3
2E_p} \left [ a\,\theta (t_2 -t_1) \,    {\cal U}_1^{\sigma\,(1)}
({\bf p}) \otimes \overline {\cal U}_1^{\sigma\,(1)} ({\bf p})
e^{ip\cdot x}+\right .\nonumber\\ &&\left.  \qquad\qquad+b\,\theta (t_1
-t_2) \, {\cal V}_1^{\sigma\,(1)} ({\bf p}) \otimes \overline {\cal
V}_1^{\sigma\,(1)} ({\bf p}) e^{-ip\cdot x} \right ] +\nonumber\\ &+&
\left [ \gamma_{\mu\nu} \partial_\mu \partial_\nu + m^2 \right  ] \int
\frac{d^3 {\bf p}}{(2\pi)^3 2E_p} \left [ a\,\theta (t_2 -t_1) \, {\cal
 U}_2^{\sigma\,(1)} ({\bf p}) \otimes \overline {\cal U}_2^{\sigma\,(1)}
({\bf p}) e^{ip\cdot x}+ \right. \nonumber\\ &&\left.
\qquad\qquad+b\,\theta (t_1 -t_2) \, {\cal V}_2^{\sigma\,(1)} ({\bf p})
\otimes \overline {\cal V}_2^{\sigma\,(1)} ({\bf p}) e^{-ip\cdot x}\right
] +\nonumber\\ &+&\left [ \widetilde \gamma_{\mu\nu} \partial_\mu
\partial_\nu + m^2 \right ] \int \frac{d^3 {\bf p}}{(2\pi)^3 2E_p} \left [
a\,\theta (t_2 -t_1) \, {\cal U}_1^{\sigma\,(2)} ({\bf p}) \otimes
\overline {\cal U}_1^{\sigma\,(2)} ({\bf p}) e^{ip\cdot x}+
\right.\nonumber\\ &&\left.  \qquad\qquad+b\,\theta (t_1 -t_2) \, {\cal
V}_1^{\sigma\,(2)} ({\bf p}) \otimes \overline {\cal V}_1^{\sigma\,(2)}
({\bf p}) e^{-ip\cdot x} \right ] +\nonumber\\ &+&\left [\widetilde
\gamma_{\mu\nu} \partial_\mu \partial_\nu - m^2 \right  ] \int \frac{d^3
{\bf p}}{(2\pi)^3 2E_p} \left [ a\,\theta (t_2 -t_1) \, {\cal
U}_2^{\sigma\,(2)} ({\bf p}) \otimes \overline {\cal U}_2^{\sigma\,(2)}
({\bf p}) e^{ip\cdot x} +\right.\nonumber\\ &&\left.
\qquad\qquad+b\,\theta (t_1 -t_2) \, {\cal V}_2^{\sigma\,(2)} ({\bf p})
\otimes \overline {\cal V}_2^{\sigma\,(2)} ({\bf p}) e^{-ip\cdot x} \right
] = \delta^{(4)} (x_2 -x_1) \end{eqnarray} can be satisfied by the
definite choice of the constant $a$ and $b$.  In the process of
calculations  I assume that the set of the analogs of the ``Pauli spinors"
in the $(1,0)$ or $(0,1)$ spaces  is a complete set in mathematical sense
and it is normalized to $\delta_{\sigma\sigma^\prime}$\,. Simple
calculations yield
\begin{eqnarray} \lefteqn{\partial_\mu \partial_\nu
\left [ a\, \theta (t_2 -t_1)\, e^{ip\cdot (x_2 -x_1)} + b\, \theta (t_1
-t_2)\, e^{-ip(x_2 -x_1)} \right ]=\nonumber}\\ &=& - \left [ a\, p_\mu
p_\nu \theta (t_2 - t_1) \exp \left [ ip\cdot (x_2 -x_1)\right ] + b\,
p_\mu p_\nu \theta (t_1 -t_2) \exp \left [ -ip \cdot (x_2 -x_1) \right ]
\right ] + \nonumber\\ &+& a\left [ - \delta_{\mu 4} \delta_{\nu 4}
\delta^{\,\,\prime} (t_2 -t_1) +i (p_\mu \delta_{\nu 4} +p_\nu \delta_{\mu
4}) \delta (t_2 -t_1) \right ] \exp \left [i {\bf p} ({\bf x}_2 - {\bf
x}_1)\right ] +\nonumber\\ &+& b\, \left [ \delta_{\mu 4} \delta_{\nu 4}
\delta^{\,\,\prime} (t_2 -t_1) + i (p_\mu \delta_{\nu 4} +p_\nu
\delta_{\mu 4}) \delta (t_2 -t_1) \right ] \exp \left [-i{\bf p} ({\bf x}_2
- {\bf x}_1)\right ] \quad; \end{eqnarray} and \begin{eqnarray} {\cal
U}_1^{(1)}\overline {\cal U}_1^{(1)} ={1\over 2} \pmatrix{m^2 & S_p
\otimes S_p\cr \overline S_p \otimes \overline S_p &m^2\cr}\quad,\quad
{\cal U}_2^{(1)}\overline {\cal U}_2^{(1)} = {1\over 2}\pmatrix{-m^2 & S_p
\otimes S_p\cr \overline S_p \otimes \overline S_p &-m^2\cr}\quad,
\end{eqnarray}
\begin{eqnarray} {\cal U}_1^{(2)}\overline {\cal U}_1^{(2)}
={1\over 2} \pmatrix{-m^2 & \overline S_p \otimes \overline S_p\cr S_p
 \otimes  S_p &-m^2\cr}\quad,\quad {\cal U}_2^{(2)}\overline {\cal
 U}_2^{(2)} = {1\over 2} \pmatrix{m^2 & \overline S_p \otimes \overline
S_p\cr S_p \otimes  S_p &m^2\cr}\quad, \end{eqnarray}
where
\begin{eqnarray} S_p &=& m + ({\bf J}\cdot {\bf p}) +\frac{({\bf J}
\cdot{\bf p})^2}{E_p+m}\quad,\\
\overline S_p &=& m - ({\bf J} \cdot{\bf p}) +
\frac{({\bf J}\cdot {\bf p})^2}{E_p+m}\quad.
\end{eqnarray}
Due to
$$\left [E_p - ({\bf J}\cdot {\bf p})\right ]  S_p \otimes S_p = m^2 \left
[ E_p + ({\bf J}\cdot {\bf p})\right ]\quad,$$ $$\left [E_p + ({\bf
J}\cdot {\bf p})\right ] \overline S_p \otimes \overline S_p = m^2 \left [
E_p - ({\bf J} \cdot{\bf p})\right ]\quad$$ after simplifying the left
side of (\ref{prop}) and comparing it with the right side we find
the constants to be equal to  $a=b=1/ 4im^2$.  Thus, if
consider all four equations (\ref{w1},\ref{w2},\ref{w11},\ref{w21})
one can use the ``Wick's formula" for the
time-ordered particle operators to find propagators:
\begin{mathletters}
\begin{eqnarray}\label{propa1}
S_F^{(1)} ( p ) &=& \frac{i\left [\gamma_{\mu\nu} p_\mu p_\nu   -  m^2
\right ]}{(2\pi)^4  (p^2  +m^2 -i\epsilon)}
\quad,\\
\label{propa2}
S_F^{(2)} ( p ) &=& \frac{i\left [ \gamma_{\mu\nu} p_\mu p_\nu   +  m^2
\right ]}{(2\pi)^4  (p^2  +m^2 -i\epsilon)} \quad,\\
\label{propa3}
S_F^{(3)} ( p ) &=& \frac{i\left [
\widetilde\gamma_{\mu\nu} p_\mu p_\nu   +  m^2
\right ]}{(2\pi)^4  (p^2  +m^2 -i\epsilon)} \quad,\\
\label{propa4}
S_F^{(4)} ( p ) &=& \frac{i\left [ \widetilde \gamma_{\mu\nu} p_\mu p_\nu
- m^2  \right ]}{(2\pi)^4  (p^2  +m^2 -i\epsilon)}\quad.  \end{eqnarray}
\end{mathletters}
The conclusion is that the states described by the equations
(\ref{w1},\ref{w2},\ref{w11},\ref{w21}) cannot propagate separately each
other, what is a principal difference comparing with the Dirac case.

Furthermore, I am able to recast the $j=1$ Tucker-Hammer equation
(\ref{eq:Tucker}) which is free of tachyonic solutions, or the Proca
equation (\ref{eq:eq}), to the form
\begin{eqnarray}\label{ME0} m^2 E_i &=& - {\partial^2 E_i \over \partial t^2} +\epsilon_{ijk} {\partial \over
\partial x_j} {\partial B_k \over \partial t} + {\partial \over \partial
x_i} {\partial E_j \over \partial x_j}\quad,\\ \label{ME1} m^2 B_i &=&
\epsilon_{ijk} {\partial \over \partial x_j} {\partial E_k
\over \partial t} +  {\partial^2 B_i \over \partial x_j^2} -{\partial
\over \partial x_i} {\partial B_j \over \partial x_j}\quad.
\end{eqnarray}
The Klein-Gordon equation (the D'Alembert equation in the massless limit)
\begin{equation}\label{DA}
\left (\frac{\partial^2}{\partial t^2} -
\frac{\partial^2}{\partial x_i^2}\right ) F_{\mu\nu} = - m^2 F_{\mu\nu}
\end{equation}
is implied ($c=\hbar=1$). Introducing vector operators one can
write equations in the following form:
\begin{mathletters}
\begin{eqnarray}\label{MY1}
{\partial \over \partial t}\, \mbox{curl}\, {\bf B} + \mbox{grad}\,
\mbox{div}\, {\bf E} - {\partial^2 {\bf E} \over \partial t^2} &=& m^2
{\bf E}\quad,\\ \label{MY2}
{\bbox \nabla}^2 {\bf B} -\mbox{grad}\, \mbox{div}\, {\bf
B} + {\partial \over \partial t}\, \mbox{curl}\, {\bf E} &=& m^2 {\bf
B}\quad.
\end{eqnarray}
\end{mathletters}
Taking into account the definitions:
\begin{mathletters}
\begin{eqnarray}
\rho_e &=&\mbox{div} \, {\bf E}\quad, \quad
{\bf J}_e = \mbox{curl} \, {\bf B} -
{\partial {\bf E} \over \partial t} \quad, \label{ME11}\\
\rho_m &=& \mbox{div} \,{\bf B} \quad, \quad {\bf J}_m = - {\partial
{\bf B} \over \partial t} - \mbox{curl}\, {\bf E} \quad,\label{ME12}
\end{eqnarray}
\end{mathletters}
the relations of the vector algebra (${\bf X}$ is an  arbitrary vector):
\begin{equation}
\mbox{curl}\, \mbox{curl}\, {\bf X} = \mbox{grad}\, \mbox{div}\, {\bf X}
-{\bbox \nabla}^2 {\bf X}\quad,
\end{equation}
and the Klein-Gordon equation (\ref{DA}) one obtains two equivalent sets
of equations, which complete the Maxwell's set of equations.
The first one is
\begin{mathletters}
\begin{eqnarray}
&& {\partial {\bf J}_e \over \partial t} +\mbox{grad} \,\rho_e
= m^2  {\bf E}\quad,\label{mm1}\\
&& {\partial {\bf J}_m \over \partial t} +\mbox{grad} \, \rho_m
=0\quad;\label{mm2}
\end{eqnarray}
\end{mathletters}
and the second one is
\begin{mathletters}
\begin{eqnarray}
&&\mbox{curl}\, {\bf J}_m =0\label{mm3}\\
&&\mbox{curl} \,{\bf J}_e = -m^2 {\bf B}\quad.\label{mm4}
\end{eqnarray}
\end{mathletters}
I would like to remind that the Weinberg equations (and, hence, the
equations (\ref{mm1}-\ref{mm4})\footnote{Beginning from the dual massive
equations (\ref{eq:Tucker2},\ref{eq:eqd}) and setting ${\bf C} \equiv
{\bf E}$, \, ${\bf D} \equiv {\bf B}$ one could obtain
\begin{mathletters}
\begin{eqnarray}
&&{\partial {\bf J}_e \over \partial t} +\mbox{grad} \,\rho_e = 0\quad,\\
&& {\partial {\bf J}_m \over \partial t}+\mbox{grad} \,\rho_m =m^2 {\bf
B}\quad; \end{eqnarray} \end{mathletters}
and
\begin{mathletters}
\begin{eqnarray} && \mbox{curl}\, {\bf J}_e =0\quad,\\ &&\mbox{curl}\,
{\bf J}_m = m^2 {\bf E}\quad.
\end{eqnarray}
\end{mathletters}
This signifies that  the
physical content spanned by massive dual fields can be different. The
reader can easily reveal  parity-conjugated equations from Eqs.
(\ref{w11},\ref{w21}).}) can be obtained on the basis of a very few number
of postulates; in fact, by using the Lorentz transformation rules for the
Weinberg bivector (or for the antisymmetric tensor field) and the
Ryder-Burgard relation~\cite{DVA1,DVA3}.

In a massless limit the situation is different. Firstly,
the set of equations (\ref{ME12}), with the left side are chosen to
be zero, is ``an identity satisfied by certain space-time derivatives
of $F_{\mu\nu}$\ldots, namely, refs.~\cite{DYSON,TANIMU}
\begin{equation}
{\partial F_{\mu\nu} \over \partial x^\sigma} +
{\partial F_{\nu\sigma} \over \partial x^\mu} +
{\partial F_{\sigma\mu} \over \partial x^\nu} = 0\quad."
\end{equation}
I belive that a similar consideration for the dual field $\widetilde
F_{\mu\nu}$ as in refs.~\cite{DYSON,TANIMU} can reveal that the same is
true for the first equations (\ref{ME11}). So, in the massless case we met
with the problem of interpretation of the charge and the current.

Secondly, in order to  satisfy the massless equations (\ref{mm3},\ref{mm4})
one should assume that the currents are represented in gradient forms
of some scalar fields $\chi_{e,m}$. What physical significance
should be attached to these
chi-functions?  In a massless case the charge densities are then (see
equations (\ref{mm1},\ref{mm2})) \begin{equation} \rho_e = -
\frac{\partial \chi_e}{\partial t} + const \quad, \quad \rho_m = -
\frac{\partial\chi_m}{\partial t} + const \quad, \end{equation} what tells
us that $\rho_e$ and $\rho_m$ are constants provided that the primary
functions $\chi_{e,m}$ are linear functions in time (decreasing or
increasing?).  One can obtain
the Maxwell's free-space equations,
in the definite choice  of the  $\chi_e$ and $\chi_m$, namely,
in the case when they are constants.

It is useful to compare the resulting equations for
$\rho_{e,m}$ and ${\bf J}_{e,m}$ and the fact of appearance of
functions $\chi_{e,m}$ with alternative formulations of electromagnetic
theory discussed in the Section I.
I belive, this concept can also be useful in analyses of the $E=0$
solutions in higher-spin
relativistic wave equations~\cite{OPPEN,MAJOR1,DVA01,DVA02}, which have
been ``baptized" by Moshinsky and Del Sol in~\cite{MOSHIN} as
the `relativistic cockroach nest'. Finally,  in ref.~\cite{GERSTEN} it was
mentioned that solutions of Eqs.  (4.21,4.22) of ref.~[71b],
see the same equations (\ref{mel},\ref{mer}) in this article, satisfy the
equations of the type (\ref{ME0},\ref{ME1}), {\it ``but not always vice
versa".} An interpretation of this statement and investigations of Eq.
(\ref{eq:Tucker})  with other initial and boundary conditions (or of the
functions $\chi$) deserve further elaboration (both theoretical and
experimental).

\medskip

\section{Conclusion}

As a conclusion of a serie of papers of mine in various journals (here I
present only a part of these investigations), one can state that due to
the present-day experimental situation the standard model seems to be able
to describe a restricted class of phenomena only.  Probably, origins of
these limitations are in methodological failures which were brought as a
result of the unreasonable (and everywhere) application of the principles
which the Maxwell's electromagnetic theory is based on.  In my opinion, it
is a particular case only.  Generalized models discussed in this paper
appear to be suitable candidates to begin to work out the unified field
theory.  All they are connected each other and it seems to indicate at the
same physical reality. I believe the Weinberg $2(2j+1)$ component
formalism  is the most convenient way for understanding the nature of
higher spin particles, the structure of the space-time and for describing
many processes with particles of the spin 0 and 1, because this formalism
is on an equal footing with the well-developed Dirac formalism for
spin-1/2 charged particles and manifests explicitly those symmetry
properties which are related with the Poincar\`e group and the group of
inversion operations. The problems in Majorana-like constructs in both
the $(1/2,0)\oplus (0,1/2)$ representation and higher-spin representations
still deserve further elaboration.

\acknowledgments

Discussions with Profs.  D.  V.  Ahluwalia, A.  E.  Chubykalo, M.  W.
Evans, A. F.  Pashkov and S.  Roy, useful suggestions of Profs.
L.  Horwitz and M.  Sachs, and critics of Prof. E. Comay as well were
invaluable for writing the paper.  I greatly appreciate many private
communications and preprints from collegues over the world.

I am grateful to Zacatecas University for a professorship.
This work has been supported in part by el Mexican Sistema
Nacional de Investigadores, el Programa de Apoyo a la Carrera Docente
and by the CONACyT, M\'exico under the research project 0270P-E.

\end{document}